# Examining graph neural networks for crystal structures: limitations and opportunities for capturing periodicity

Sheng Gong[1*], Tian Xie[2], Yang Shao-Horn[1,3], Rafael Gomez-Bombarelli[1], and Jeffrey C. Grossman[1*]

[1]Department of Materials Science and Engineering, Massachusetts Institute of Technology, MA 02139, USA
[2]Microsoft Research, Cambridge, United Kingdom CB1 2FB
[3]Department of Mechanical Engineering, Massachusetts Institute of Technology, MA 02139, USA

# Abstract

Historically, materials informatics has relied on human-designed descriptors of materials structures. In recent years, graph neural networks (GNNs) have been proposed for learning representations of crystal structures from data end-to-end producing vectorial embeddings that are optimized for downstream prediction tasks. However, a systematic scheme is lacking to analyze and understand the limits of GNNs for capturing crystal structures. In this work, we propose to use human-designed descriptors as a bank of human knowledge to test whether black-box GNNs can capture the knowledge of crystal structures. We find that current state-of-the-art GNNs cannot capture the periodicity of crystal structures well, and we analyze the limitations of the GNN models that result in this failure from three aspects: local expressive power, long-range information, and readout function. We propose an initial solution, hybridizing descriptors with GNNs, to improve the prediction of GNNs for materials properties, especially phonon internal energy and heat capacity with 90% lower errors, and we analyze the mechanisms for the improved prediction. All the analysis can be extended easily to other deep representation learning models, human-designed descriptors, and systems such as molecules and amorphous materials.

# Introduction

Recently, machine learning (ML) has been widely employed to predict properties of materials[1-8]. Conversion of crystal structures into machine-readable numerical representations is one of the most critical steps for applications of ML in materials science[9]. In general, there are two approaches to convert crystal structures into numbers: human-designed description and deep representation learning.

Human-designed descriptors are based on people's understanding of compositions and structures of materials, and therefore, one can easily understand the meaning of such descriptors[10]. Mean electronegativity and difference of atomic radius of elements in materials are examples of compositional descriptors, and mean bond length and difference of coordination number of atoms in crystal structures are examples of structural descriptors of materials. Beyond simple descriptors, researchers have recently proposed a series of more complex descriptors for materials, such as Magpie[11] compositional descriptors, classical force-field inspired descriptors (CFID)[12], Coulomb matrix[13], and fragment descriptors[14]. Although ML models based on human-designed descriptors have achieved some success in revealing the trend between human-understandable characteristics of materials and properties, by definition these descriptors contain only known information. Consequently, employing only these descriptors to learn and predict materials properties could miss key structure-property relationships that are currently unknown.

Deep representation learning refers to ML models that learn the numerical representation of materials automatically during the training of the ML models. Although the learned representations are generally less understandable compared with human-designed descriptors, deep representation learning can uncover unknown patterns of structure-property relationships. Since materials can be intuitively represented as graphs, with atoms forming the nodes and bonds forming the edges, graph neural networks (GNNs) have become the state-of-the-art deep representation learning method for materials science. SchNet[15] and CGCNN[16] are two classic GNN architectures designed for materials.

They update the representations of each atom by the types of neighboring atoms and the bond length between atoms, and pool all the updated atom representations into an overall representation of the structure. In later variants of GNN for materials such as iCGCNN[17], MEGNet[18] and GATGNN[19], bond representations are also updated during the convolution. Through multiple layers of graph convolutions, these models can implicitly encode many-body interactions. To explicitly encode many-body interactions, Gasteiger *et al.* proposed DimeNet[20] and GemNet[21] for molecules, and Choudhary *et al.* proposed ALIGNN for periodic crystal structures[22], where atom representations (one-body), bond representations (two-body) and bond angle representations (three-body) are all updated during the convolution via the construction of line graph (the nodes of the line graph are edges in the original graph, and edges of the line graph are angles between edges in the original graph). Together with other studies using higher-order information to improve the expressiveness of GNN[23, 24], ALIGNN-$d$[25], a recent variant of ALIGNN, updates the dihedral angle representation (four-body) by constructing line graph of line graph. Very recently, Batatia *et al.*[26] proposed a general formalism to encode local atomic environments by GNN with arbitrary body-order. Other efforts have also been made to improve GNN for crystal structures such as the inclusion of state attributes in MEGNet[18], attention mechanism in GATGNN[19], representations equivariant to rotations and inversion in E3NN[27, 28], the use of structure motifs in AMDNet[29], prediction of tensorial properties in ETGNN[30], and exploitation of correlations in spectral properties in Mat2Spec[31].

Although these variants of GNNs have achieved successes in learning materials properties, for capturing crystal structures the improvements are mainly based on human intuition of the local bonding environment, such as explicitly encoding bond angle (three-body interaction) and dihedral angle (four-body) information, representations equivariant to rotations (orientations of bond vectors), and structure motifs. In general, prediction of materials properties is still challenging[22], and there is still no systematic approach and quantitative metric to analyze and understand the limitations of GNNs for crystal structures, especially for global information of crystal structures beyond local atomic environments.

In this work, we propose a systematic approach to analyze and quantify the limitations of GNNs for crystal structures, and propose a way to improve the GNN models for predicting materials properties. As illustrated in Figure 1, we use human-designed descriptors as a bank of knowledge to test whether the current GNN models can capture certain knowledge of crystal structures. We test the GNNs by employing them to learn and predict the human-designed descriptors, and use the prediction accuracy as a quantitative metric for evaluation. The underlying assumption is that, if the model can accurately predict the descriptor, then the model can capture the knowledge behind the descriptor, otherwise the model may not be able to capture certain pieces of information about crystal structures. We find that the GNNs do not capture the periodicity of crystal structures well, and we analyze the reasons for this failure in some detail. We further hybridize the deep learning models with the human-designed descriptors, and test the descriptors-hybridized models on a range of important materials properties. We find that hybridization of GNNs and descriptors can result in up to 90% decrease of errors for predictions of phonon-related properties compared with original GNNs.

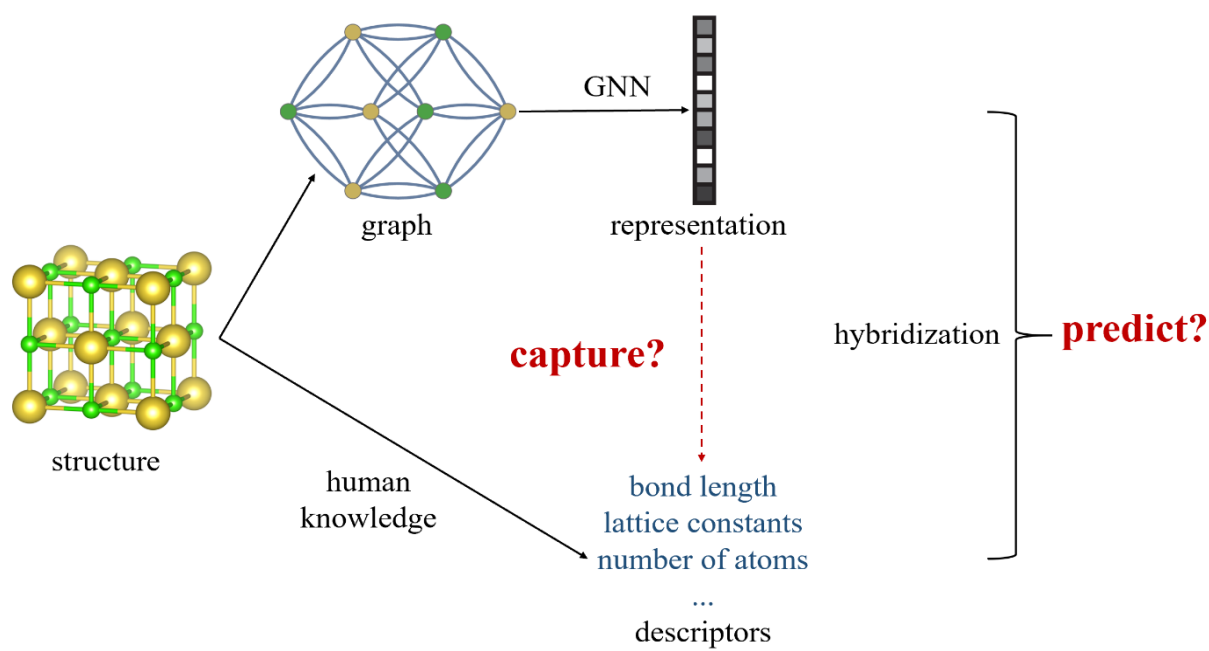

**Figure 1.** Schematic of analyzing whether a GNN can capture knowledge of crystal structures behind human-designed descriptors, and whether hybridization of GNN and human-designed descriptors can improve prediction performance for materials properties.

## Results

**Brief review of CGCNN and ALIGNN.** In this work, we choose CGCNN and ALIGNN as two examples of GNNs to investigate their ability to capture human-designed descriptors, and as examples for improving prediction ability by hybridization with descriptors. CGCNN is one of the classic and most frequently used GNNs for materials, while ALIGNN is one of the state-of-the-art models for prediction of materials properties with the best performance on its in-house test set[22] and the open Matbench test set[32]. Both CGCNN and ALIGNN are specifically designed for predicting properties of periodic materials and have well-documented open-source codes to use and adapt. CGCNN explicitly encodes two-body interactions, and ALIGNN explicitly encodes three-body interactions. Although there are already GNNs that explicitly encode *n*-body interactions ($n \geq 4$)[23-26], they are not specifically designed for prediction of properties of periodic crystal structures or lack a comprehensive benchmark yet, and are thus not examined in this work.

The architecture of CGCNN (https://github.com/txie-93/cgcnn) is summarized in equations (1) to (3):

$$a_i^{(n+1)} = a_i^{(n)} + \sum_{j,k} \sigma(m_{(i,j)_k}^{(n)} \boldsymbol{W}_{gate}^{(n)}) \odot g(m_{(i,j)_k}^{(n)} \boldsymbol{W}_{message}^{(n)}) \ldots\ldots (1),$$

$$m_{(i,j)_k}^{(n)} = a_i^{(n)} \oplus a_j^{(n)} \oplus b_{(i,j)_k} \ldots\ldots (2),$$

$$\text{Output} = \text{AGG}(a_1^{(n^*)}, a_2^{(n^*)}, \ldots, a_N^{(n^*)}) \ldots\ldots (3).$$

Here, $a_i^{(n)}$ denotes the representation of atom *i* at layer *n*, $b_{(i,j)_k}$ representation of the $k^{\text{th}}$ bond between atom *i* and *j* at layer *n*, $n^*$ the final convolution layer, $\boldsymbol{W}_{gate}^{(n)}$ the gate matrix at layer *n*, $\boldsymbol{W}_{message}^{(n)}$ the message matrix at layer $n$, $m_{(i,j)_k}^{(n)}$ the message from atom $j$ to atom $i$ via the $k^{th}$ bond, $\odot$ element-wise multiplication, $\oplus$ concatenation, $\sigma$ and $g$ non-linear activation functions, AGG the aggregation (readout) function. In CGCNN, the implemented aggregation function can be written as:

$$\text{Output} = \text{FCN}(\frac{1}{N_a}\sum_{i=1}^{N_a} a_i^{n^*}) \text{ ...... (4)},$$

where the output is calculated by first taking the average of all atom representations, then feeding the averaged representation to a fully connected network. The reason for using average pooling (equation (4)) is that intensive materials properties, such as band gap and refractive index, are invariant to the (supercell) size of crystal structures. In summary, in each convolution layer, CGCNN uses neighboring atoms and bond length as messages to each atom, and updates each atom representation by feeding the messages into a gate layer and a message processing layer. After convolutions, CGCNN pools all atom representations by taking the average and input the pooled material representation into a fully connected network to compute the property.

The architecture of ALIGNN (https://github.com/usnistgov/alignn) is summarized in equations (5) to (10), with equations (5) to (7) describing the atomistic graph, and equations (8) to (10) the line graph:

$$a_i^{(n+1)} = a_i^{(n)} + g(a_i^{(n)}\boldsymbol{W}_{self}^{(n)} + \sum_{j,k} g'(b_{(i,j)_k}^{(n)})\boldsymbol{W}_{message}^{(n)} a_j^{(n)}) \text{ ...... (5)},$$

$$b_{(i,j)_k}^{(n+1)} = b_{(i,j)_k}^{(n)} + g(m_{(i,j)_k}^{(n)}\boldsymbol{W}_{gate}^{(n)}) \text{ ...... (6)},$$

$$m_{(i,j)_k}^{(n)} = a_i^{(n)} \oplus a_j^{(n)} \oplus b_{(i,j)_k}^{(n)} \text{ ...... (7)},$$

$$b_i^{(n+1)} = b_i^{(n)} + g(b_i^{(n)}\boldsymbol{W'}_{self}^{(n)} + \sum_{j,k} g'(t_{(i,j)_k}^{(n)})\boldsymbol{W'}_{message}^{(n)} b_j^{(n)}) \text{ ...... (8)},$$

$$t_{(i,j)_k}^{(n+1)} = t_{(i,j)_k}^{(n)} + g(m'^{(n)}_{(i,j)_k}\boldsymbol{W'}_{gate}^{(n)}) \text{ ...... (9)},$$

$$m'^{(n)}_{(i,j)_k} = b_i^{(n)} \oplus b_j^{(n)} \oplus t_{(i,j)_k}^{(n)} \text{ ...... (10)}.$$

Here, $t$ denotes the representation of bond angle, and other symbols share similar meaning to that of CGCNN. In summary, in each convolution layer, ALIGNN updates atom representations by neighboring atoms and bonds, updates bond representations twice: by connected atoms, and by neighboring bonds and bond angles, and updates bond angle representations by connected bonds. After convolutions, ALIGNN uses average pooling in equation (4) to collect atom representations as the material representation, and calculates the property by a fully connected network.

For building periodic crystal graphs, in their default settings, both CGCNN and ALIGNN use a cut-off radius of 8 Å and maximum 12 nearest neighbors, and both of them use radial basis functions to expand the interatomic distances for initialization of bond representations. ALIGNN also uses radial basis functions to expand cosines of bond angles for initialization of bond angle representations. CGCNN updates atom features by 3 graph convolution layers, and ALIGNN updates atom features by 4 line graph convolution layers (equations (5) to (10)) and 4 normal graph convolution layers (equations (5) to (7)). In the following, we use CGCNN and ALIGNN with the default setting unless otherwise specified.

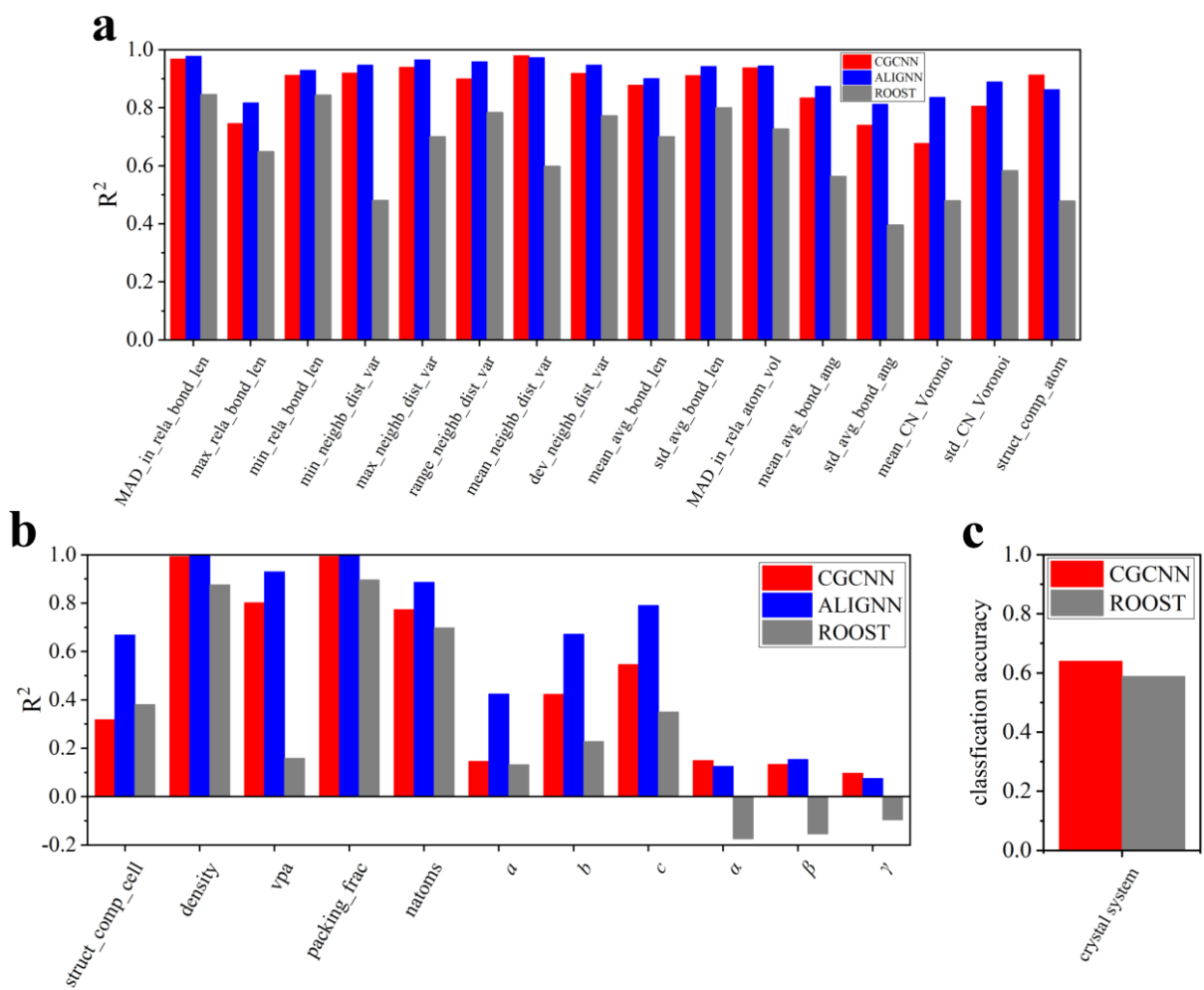

**Figure 2. Learning and predicting human-designed descriptors to examine whether the GNNs can capture certain human knowledge. a** and **b** $R^2$ scores of predictions of human-designed structural descriptors from CGCNN, ALIGNN and ROOST for local and global structural descriptors, respectively. The full names of the descriptors are listed in Table 3. **c** Classification accuracy of crystal system of crystal structures from CGCNN and ROOST.

**Learning and predicting human-designed descriptors.** In this section, we employ CGCNN and ALIGNN to learn and predict structural descriptors of a subset of crystal structures in the Materials Project database[33] ("MP dataset" as below; details in the **Methods** section) to examine the ability of the GNNs to capture certain knowledge behind the descriptors. As a baseline, we also use ROOST[34], one of the most powerful composition-only deep learning models, to learn and predict the structural descriptors.

In Figure 2a, we show the accuracies of predictions of some of the most basic local structural descriptors calculated by matminer[35] from CGCNN, ALIGNN, and ROOST in terms of $R^2$ scores ($R^2 = 1 - \frac{\Sigma(y_i - y_{i,true})^2}{\Sigma(y_{i,true} - \bar{y})^2}$, $y_i$ predicted value, $y_{i,true}$ true value, $\bar{y}$ mean of true values). We can see that, for most local structural descriptors, both CGCNN and ALIGNN can properly predict them with $R^2$ scores close to or higher than 0.8, and both of the two structure-based models outperform the composition-only model (ROOST). Because local descriptors in this work are essentially statistics of local environments around each atom, the explicit encoding of bond angles (three-body interaction) in ALIGNN might explain why ALIGNN outperforms CGCNN for learning local structural descriptors as in Figure 2a. The cases with lower $R^2$ scores in Figure 2a, such as max_rela_bond_len (maximum relative bond length) and std_avg_bond_ang (standard deviation of average bond angles), can be attributed to the fact that average pooling (equation (4)) is used by both CGCNN and ALIGNN to obtain the mean statistics of atom representations, while the two descriptors here describe the maximum and standard deviation of a collection of atomic environments. For the coordination number defined by the Voronoi method[36], we can see that CGCNN cannot capture it very well, because the Voronoi coordination number is defined in a more complicated way than that defined by the method of nearest neighbors as in CGCNN and ALIGNN. The encoded angular information boosts the ability of ALIGNN to capture coordination number, as the algorithm to determine the Voronoi coordination number heavily uses the angular information[36]. We anticipate that, GNNs that determine neighbors by the Voronoi method, such as the iCGCNN[17], might have strong ability to capture the Voronoi coordination number.

In addition to basic local descriptors, we also test the ability of CGCNN and ALIGNN to capture knowledge behind more global structural descriptors. In Figure 2b, we show the accuracies of predictions of some of the most basic global structural descriptors calculated by matminer[35] and pymatgen[37] from CGCNN, ALIGNN, and ROOST. Both CGCNN and ALIGNN can predict density, vpa (volume per atom), packing fraction, and natoms (number of atoms in the primitive cell; in this work, the "primitive cell" is defined as the Niggli reduced cell[38, 39]) with $R^2$ scores close to or higher than 0.8. Note that the Niggli reduced cell is unique for a given structure, and more discussions are provide in the Supporting Information. However, they cannot predict struct_comp_cell (structural complexity per cell[40]) and lattice constants ($a, b, c, \alpha, \beta, \gamma$; in this work, $a$ denotes the length of the longest lattice vector, $c$ the shortest, and $\alpha$ denotes the largest lattice angle, $\gamma$ the smallest) well. Both structure-based models outperform the composition-only model, and ALIGNN outperforms CGCNN, except for $\alpha$ and $\gamma$. Note that the poor predictions of lattice constants and accurate predictions of density and vpa (volume per atom) do not contradict with each other, and more discussions are provided in the Supporting Information.

In Figure 2c, we show the accuracy of multi-class classification of crystal system of crystal structures from CGCNN and ROOST (we do not employ ALIGNN as its whole framework, from data collection to model output, is structured with the style of binary-classification), from which we can see that, although CGCNN can classify crystal system of structures with around 60% accuracy, ROOST can also achieve similar accuracy, which agrees with CRYSPNet that the type of crystal structure of inorganic materials can be partially determined by composition[41]. The similar classification ability of CGCNN and ROOST towards crystal system suggests that, the geometric information encoded by CGCNN does not significantly help to classify crystal system, which supports our conclusion that GNN cannot capture lattice constants well, as crystal system is determined by the relationship between lattice constants.

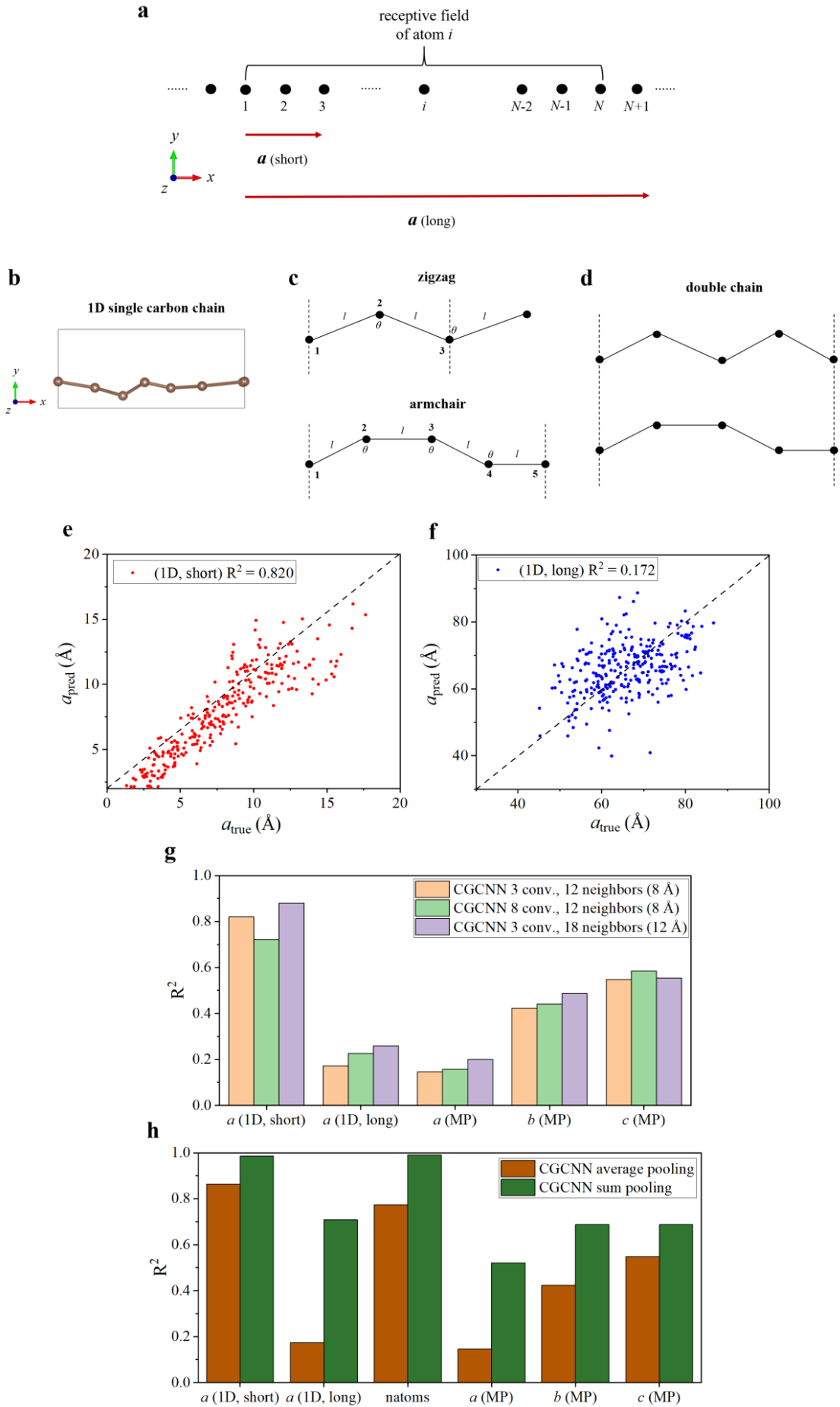

**Figure 3. Limitations of GNNs for capturing periodicity. a** Illustration of the receptive field of an atom in a GNN and the periodicity of a 1-dimensional (1D) structure. Here, atom $i$ receives information from atoms 1 to $N$, and two cases of periodicity are plotted: the short periodicity from atom 1 to 3 and the long periodicity from atom 1 to $N+1$. **b** Illustration of 1D single carbon chains as toy structures. The chains are along the $x$ direction with periodicity, with random displacement of each atom in the $y$ and $z$ directions. **c** Illustration of 1D chains with zigzag and armchair configuration, respectively. **d** Illustration of 1D double chain. **e** and **f** $a_{\text{true}}$ versus $a_{\text{pred}}$ of the datasets of 1D short chains and 1D long chains from default CGCNN, respectively. **g** $R^2$ scores of predictions of $a$ of 1D short chains and 1D long chains, and $a, b, c$ of the MP dataset, from default CGCNN, CGCNN with 8 convolution layers, and CGCNN connecting 18 nearest neighbors within 12 Å, respectively. **h** $R^2$ scores of prediction of $a$ of 1D short chains and 1D long chains, and natoms, $a, b, c$ of the MP dataset from CGCNN with average pooling and CGCNN with sum pooling, respectively.

**Limitations of GNNs for capturing periodicity.** Although previous works have suggested that lattice constants of crystal structures are learnable based on only compositions[41, 42], the results in this work show that even with structures as input, CGCNN and ALIGNN cannot capture lattice constants well. In this section, we analyze the possible reasons for the failure and obtain insights for improving GNNs for crystal structures.

Lattice constants describe the periodicity of atomic structures. If $A(r)$ describes the type of atom at position $r$ ("none" if there is no atom at that position), and if $R$ is a linear combination of lattice vectors, then periodicity requires that:

$$A(r) = A(r + R)......(11).$$

In 3-dimensional (3D) space, we need 3 linearly independent lattice vectors to describe the periodicity of atomic structures. Lattice constants describe the periodicity by the length of lattice vectors $(a, b, c)$ and angles between lattice vectors $(\alpha, \beta, \gamma)$. To simplify the analysis, in addition to 3D crystal structures in the MP dataset, we also consider the toy cases of quasi-1D atomic chains as in Figure 3a, where periodicity is imposed only along the $x$ direction. In this quasi-1D space, we only need the length of the lattice vector ($a$) to describe the periodicity: $A(r) = A(r + a)$.

For GNNs with average pooling in equation (4), since they use the average local atomic environments to represent the atomic structures, they capture periodicity by learning how equation (11) affects the local atomic environments within the receptive fields of atoms in the GNNs. The receptive field of each atom describes the range of the space where information can be propagated to the atom through the GNNs, and it depends on the number of neighbors each atom can connect to and the number of convolution layers in the GNNs:

range of receptive field $\propto$ number of neighbors $*$ number of convolutions……(12).

If the length of the periodicity (length of lattice vector) is smaller than the length of the receptive fields of atoms in the GNNs, then the GNNs might be able to capture the short periodicity; however, if the length of the periodicity is larger than the length of the receptive fields of atoms, then in principle the GNNs cannot capture the long periodicity. For example, as in Figure 3a, if the periodicity is short, such as the top red arrow which requires that atom 1 and atom 3 (atom *n* and atom *n*+2) have the same type and coordinates in the *y* and *z* directions, then the local atomic environment input to atom *i* is constrained by such periodicity, and the GNNs might be able to capture the constraint and periodicity. However, if the periodicity is long, such as the bottom red arrow describing that the periodicity is imposed between atom 1 and atom *N*+1 (one atom beyond the receptive field), then there is no constraint inside the receptive field of atom *i*, and the GNNs cannot capture the long constraint and the periodicity.

To analyze the behaviors of GNNs on capturing periodicity, in this section, we introduce toy datasets of quasi-1D carbon chains as illustrated in Figure 3b ("1D dataset" as below; details in the **Methods** section), and we create two versions of the 1D datasets: a short dataset where the periodicity of each chain is shorter than the receptive fields of atoms (1D, short), and a long dataset where the periodicity is longer than the receptive fields (1D, long). We use the default CGCNN to learn and predict the length of lattice vector ($a$) of the two datasets, and in Figure 3e and 3f, we show the predicted $a$ versus true $a$ of the two datasets. We can see that, for the short chains, CGCNN can predict $a$ well with the $R^2$ score larger than 0.8, while for the long chains, CGCNN cannot predict $a$ well. The prediction results of $a$ of the quasi-1D carbon chains support our analysis above that GNNs might be able to capture short periodicity while hard to capture long periodicity. Note that, the analysis for 1D long chains can represent thousands of materials in the Materials Project dataset where the receptive field of typical GNNs might be shorter than the periodicity of the crystal structures. More discussions are provided in Figure S5 in the Supporting Information.

Although the periodicity of most short chains in this work can be learned properly as in Figure 3e, theoretically, GNNs with limited local expressive power are not able to fully determine the periodicity. Since Chen *et al.*[43] have proved the equivalence between the ability of GNNs to distinguish graphs and approximate graph functions, if a GNN cannot distinguish two atomic graphs with different periodicity, then the GNN cannot fully determine the graph function describing the periodicity. In Figure 3c, we show two cases of 1D chains: a 1D zigzag chain and a 1D armchair chain, which represent structure prototypes of some real crystal structures such as organic crystals[44] and metal chalcogenides[45]. If a GNN uses only diatomic distances to encode local atomic environments (such as CGCNN), and if the GNN only connect to the nearest neighbors (**1** to **2**), then the GNN cannot distinguish different zigzag and armchair 1D chains with the same bond length but different bond angles and cannot capture the angle dependence of $a$. If the GNN can connect to the second nearest neighbors (**1** to **3**), then the GNN is able to distinguish zigzag and armchair 1D chains with different bond angles; however, it is still not able to distinguish between zigzag and armchair chains with the same bond length and bond angle. The analysis suggests that, to improve the ability of GNNs to capture short periodicity, it might be helpful to increase the local expressive power of GNNs to distinguish structures with different periodicity.

In Figure 3g, we show the effects of number of convolution layers and number of neighbors of

CGCNN on capturing periodicity of 1D chains and 3D crystal structures. As in equation (12), both increasing number of convolution layers and increasing number of neighbors extend the receptive fields of atoms in CGCNN, and as the discussion of zigzag and armchair chains above, increasing number of neighbors can lead to higher local expressive power to distinguish graphs. From Figure 3g, we can see that for short chains, increasing the number of neighbors leads to better prediction of $a$, which supports our suggestion above that improving the local expressive power can help to capture short periodicity. For long chains, both increasing number of neighbors and number of convolution layers result in better prediction of $a$, indicating that extending the receptive fields of atoms in CGCNN can help to capture long periodicity. As for the length of lattice vectors of real 3D structures in the MP dataset (mixed with short and long structures as in Figure S5b in the Supporting Information), we can see that both increasing number of neighbors and number of convolution layers lead to better prediction of $a, b, c$. However, we find that increasing number of convolution layers by 133% and number of neighbors by 50% just lead to moderate improvement of prediction of $a, b, c$. There are two possible reasons for the small degree of improvement: 1) as shown in equation S4 in the Supporting Information, in real 3D closely-packed materials, the radius of receptive field ($r$) is proportional to $\sqrt[3]{n}$ ($n$ is the number of neighbors). Therefore, although in Figure 3 we test CGCNN with 50% higher cut-off radius and maximum number of neighbors, for closely-packed 3D crystal structures, the ratio of elongation of receptive field might just be $\sqrt[3]{1.5} \approx 1.14$, which limits the impact of increasing cut-off radius and maximum number of neighbors on capturing periodicity. 2) even if the receptive field can be effectively elongated, the problem of over-smoothing and over-squashing[46, 47] associated with GNNs with too many convolutions and neighbors might deteriorate the expressive power of GNNs, which also limits the ability of GNNs to capture periodicity. In fact, as shown in Figure 3g, for 1D short chains, increasing the number of convolution layers deteriorates the power of CGCNN to capture periodicity, which might be attributed to over-smoothing and over-squashing[46, 47] as discussed above. Since the cost of graph convolution operations is proportional to number of convolution layers and neighbors, we suggest that simply increasing number of convolution layers and neighbors might not be an ideal way to improve the ability of GNNs to capture periodicity.

The analysis above is based on average pooling in equation (4). If we use sum pooling in equation (13) with size extensibility:

$$\text{Output} = \sum_{i=1}^{N_a} \text{FCN}(a_i^{n^*}) \quad \text{......} \quad (13),$$

then the GNNs capture periodicity by summing contributions of each atom to the lattice vectors. In Figure 3h, we show the $R^2$ scores of predictions of $a$ of the 1D chains and natoms, $a$, $b$, $c$ of the MP dataset from CGCNN with average pooling and sum pooling, respectively. For 1D short chains, sum pooling can lead to better prediction of $a$ than average pooling, which might be explained by the fact that sum pooling is more expressive than average pooling[48]. For 1D long chains, sum pooling can result in significantly better prediction of $a$ than average pooling, because average pooling requires that each atom encodes information from one end of the long primitive cell to the other end to capture the structural constraint imposed by the periodicity, while sum pooling needs only local contributions of each atom to the lattice vectors. Consistent with the results of $a$ of the 1D chains, for natoms, $a$, $b$, $c$ of the MP dataset, sum pooling can also result in better prediction than average pooling. The stronger ability of sum pooling to capture periodicity might lead to better prediction of extensive materials properties, and in Table 2 and Figure S2 in the Supporting Information, we show that sum pooling can provide better prediction than average pooling for phonon internal energy ($U$), phonon heat capacity ($C_v$) and total magnetization ($M$).

Despite the improvement, we suggest that sum pooling is not an ideal solution to the challenge of capturing periodicity. Periodicity and lattice constants of the primitive cells do not scale with supercell size and are intensive characteristics of crystal structures. In principle, sum pooling cannot be employed in machine learning of materials' intensive properties due to the requirement of (supercell) size invariance[16]. The improvement of sum pooling over average pooling in Figure 3h and S2 is based on the fact that primitive cells of crystals are used as input to the GNNs in this work. Even if only primitive cells are input to the GNNs, sum pooling might also fail to capture periodicity in some cases, as periodicity does not always scale with the number of atoms in the primitive cells. For example, in Figure 3d we show the case of 1D double chains. Compared with 1D single chains in Figure 3b and 3c, 1D double chains can have similar periodicity but twice number of atoms. In Figure S3, we show that, compared with the datasets with only 1D single chains, sum pooling is less powerful to capture the periodicity of the datasets mixed with 1D single and double chains.

From Figure 2b, we can see that ALIGNN outperforms CGCNN in the prediction of natoms, $a$, $b$, $c$ of the MP dataset. This improved predictive ability could result from two factors: on the one hand, ALIGNN has stronger local expressive power than CGCNN as it explicitly encodes bond angles, and on the other hand, ALIGNN has a larger receptive field than CGCNN, as in each convolution layer in CGCNN, a node receives messages only from the first shell of bonds and neighbors in equation (2), while in each convolution layer in ALIGNN, a node also receives messages from the second shell of bonds in equation (10). Although with the default settings ALIGNN has 8 convolution layers while CGCNN has only 3 convolution layers, from Figure 3g we can see that increasing the number of convolution layers of CGCNN to 8 leads to only moderate improvement and cannot make the predictions of $a$, $b$, $c$ from CGCNN as accurate as that of ALIGNN, which shows that the difference in the number of convolution layers in the default CGCNN and ALIGNN is not a critical factor on their relative ability to capture periodicity of the MP dataset, which might result from the issues of over-smoothing and over-squashing of deep GNNs[46, 47] as discussed above.

In Figure 2b and S4, we show that both CGCNN and ALIGNN cannot learn the lattice angles of the primitive cell well, and sum pooling, more convolutions, and more neighbors do not improve the prediction. Here we partially attribute these results to the artificial choice of lattice angles. More discussions regarding the determination of the primitive cell are provided in the Supporting Information. In this work, we choose the set of six parameters $(a, b, c, \alpha, \beta, \gamma)$ as a widely used rotationally invariant representation of lattice vectors, which might add artificial difficulty to the learning of periodicity. For example, in addition to the problems associated with learning and prediction of $a$ in the 1D cases as above, for learning and prediction of the length of the longest lattice vector of 3D structures the GNNs need to first identify which dimension is associated with the largest length, then determine the largest lattice length. For fairer evaluation, it is necessary to develop representations of periodicity that are equivariant to rotations to avoid this additional difficulty.

According to MLatticeABC[42] and CRYSPNet[41], lattice constants of high-symmetry materials are reported to be learnable based on only compositions of materials, while here we show that lattice constants are not learnable by the GNNs even with structures as input. In the previous works, materials with different symmetry are learned separately, and lattice constants of high-symmetry materials are reported to be more learnable than that of low-symmetry materials, while in this work the MP dataset is mixed with different symmetries and is biased to materials with low symmetry. More details about the MP dataset are provided in the **Methods** section.

In this section, we discuss the limitations of the GNNs on capturing periodicity mainly in three aspects: limited local expressive power, difficulty of capturing long-range information beyond receptive fields of atoms, and average pooling as the readout function. For local expressive power, advancements of GNNs to capture more structural characteristics, such as ALIGNN-$d$ for dihedral angles[25] and equivariant representations for orientation of bond vectors[27, 28], might be helpful to better capture periodicity of structures with lattice vectors shorter than the receptive fields. For long-range information, on the one hand efforts to train very deep GNNs effectively and efficiently, such as DeeperGATGNN[49], are helpful to extend the receptive fields of atoms. On the other hand, the idea of topological message passing[50, 51], periodic self-connecting[52], and propagating information in the reciprocal lattice[53] might be useful to capture long-range information by connecting nodes in the same cell complex that are far from each other, and the idea of Implicit Graph Neural Networks (IGNN)[54] might also be useful to bypass the problems associated with training very deep graph neural networks by obtaining implicitly defined state vectors from a fixed-point equilibrium equation. It is also necessary to further develop readout functions to collect the long-range information with size invariance, and the whole-graph self-attention based readout function used in GraphTrans[55] might be a good starting point to collect global information of crystals.

**Descriptors-hybridized deep representation learning.** From the results of learning human-designed descriptors, we know that GNNs might not capture all knowledge behind human-designed descriptors. One way to overcome the issue is to design better GNN architectures for specific information, such as long-range information. Another way to overcome the issue is to input the missing knowledge into the deep representation learning models. Although this idea is straightforward and used in previous works[56, 57], such as the incorporation of lattice vectors in GeoCGNN[56], the previous works did not explore the role of this added information systematically. In this section we show the mechanisms of how inputting certain knowledge to GNNs improves the prediction of materials properties, and we find that the hybridization with descriptors can lead to a significant improvement for prediction of some materials properties, especially vibrational properties that largely depend on periodicity.

We construct the descriptors-hybridized graph neural networks as below:

$$\text{Output} = \text{FCN}(\frac{1}{N_a}\sum_{i=1}^{N_a} a_i^{n^*} \oplus \text{descriptors}) \quad \ldots\ldots (14).$$

In other words, we concatenate the vector of descriptors to the vector of learned representation, and input the hybridized representation vector to the fully-connected network.

In Table 1, we show the prediction results of descriptors-hybridized CGCNN and ALIGNN (de-CGCNN and de-ALIGNN) on 13 materials properties, with the full names of the abbreviations of properties in Table 3, and detailed errors in Table S1. The set of properties includes final total energy ($E_{\text{fin.}}$), band gap ($E_g$), bulk and shear modulus ($K$ and $G$), lattice thermal conductivity ($\kappa$), phonon internal energy and heat capacity at 300 K ($U$ and $C_v$), Poisson ratio ($v$), modulus of the piezoelectric tensor ($\|e\|_\infty$), electronic and total dielectric constant ($\varepsilon_e$ and $\varepsilon_t$), refractive index ($n$) and total magnetization ($M$). The errors of the machine learning models are presented using the metric MAE/MAD = $\frac{\sum|y_i - y_{i,true}|}{\sum|y_{i,true} - \bar{y}|}$, which is invariant to scaling and used in the ALIGNN paper[22]. Typically, a model with a MAE/MAD < 0.2 is considered a good predictive model[11, 22]. We can see that de-CGCNN has improved prediction performance for most properties compared with the original CGCNN, and de-ALIGNN has close-to or larger than 10% improvement for four properties ($\kappa$, $U$, $C_v$, and $M$) and similar performance for other properties compared with the original ALIGNN. Both de-CGCNN and de-ALIGNN outperform the descriptors-only model for all properties, regardless of whether CGCNN and ALIGNN outperform the descriptors-only model.

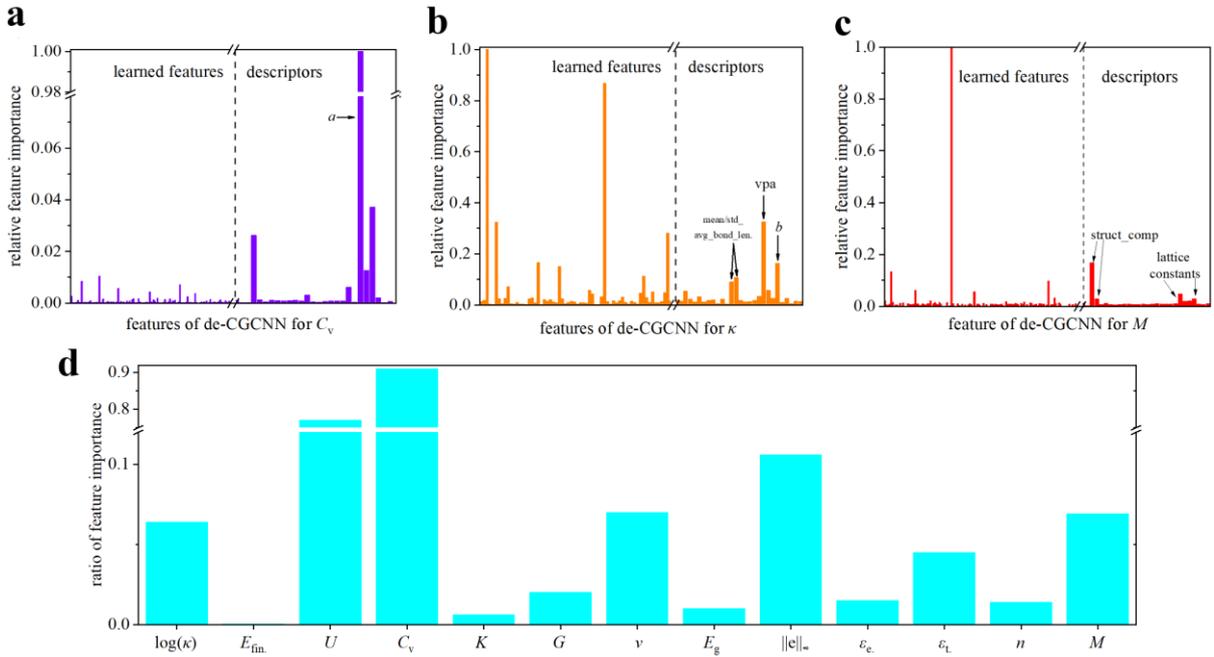

**Figure 4. Feature importance of descriptors-hybridized GNNs. a**, **b** and **c** Relative feature importance of representations from de-CGCNN for $C_v$, $\kappa$, and $M$, respectively. **d** Ratio of feature importance of input human-designed descriptors to the total feature importance from de-CGCNN for the 13 materials properties.

Table 1. Prediction results of machine learning models for 13 materials properties. The error in each cell is the MAE/MAD ratio. Here all the GNN models are used with average pooling, and comparison between CGCNN, ALIGNN and de-CGCNN, de-ALIGNN with different pooling functions and normalizations is provided in Table 2. A more detailed version of this table is provided in Table S1 in the Supporting Information.

|  | descriptors-only model | CGCNN | de-CGCNN | improvement | ALIGNN | de-ALIGNN | improvement |
|---|---|---|---|---|---|---|---|
| $\log(\kappa)$ | 0.365 | 0.347 | 0.260 | 25.1% | 0.276 | **0.248** | 10.1% |
| $E_{\text{fin.}}$ | 0.398 | 0.057 | **0.042** | 26.3% | 0.045 | 0.052 | -15.5% |
| $U$ | 0.140 | 0.670 | **0.060** | 91.0% | 0.534 | 0.101 | 81.1% |
| $C_v$ | 0.111 | 0.759 | **0.058** | 92.4% | 0.632 | 0.089 | 85.9% |
| $K$ | 0.348 | 0.217 | 0.215 | 0.92% | **0.204** | 0.206 | -0.98% |
| $G$ | 0.489 | 0.397 | 0.362 | 8.82% | 0.337 | **0.326** | 3.26% |
| $v$ | 0.800 | 0.801 | 0.788 | 1.62% | 0.716 | **0.702** | 1.95% |
| $E_g$ | 0.511 | 0.231 | 0.218 | 5.63% | **0.196** | 0.198 | -1.02% |
| $\|e\|_\infty$ | 0.793 | 0.813 | 0.762 | 6.27% | 0.796 | **0.742** | 6.78% |
| $\varepsilon_{\text{e.}}$ | 0.496 | 0.264 | 0.242 | 8.33% | **0.213** | 0.214 | -0.47% |
| $\varepsilon_{\text{t.}}$ | 0.670 | 0.563 | 0.549 | 2.49% | 0.498 | **0.489** | 1.81% |
| $n$ | 0.450 | 0.240 | 0.205 | 14.6% | 0.197 | **0.194** | 1.52% |
| $M$ | 0.622 | 0.412 | 0.340 | 17.5% | 0.340 | **0.308** | 9.41% |

Table 2. MAE of different regression settings for phonon internal energy ($U$), phonon heat capacity ($C_v$), and total magnetization ($M$). The value inside the parenthesis in each cell is the MAE/MAD ratio.

|  | Unit | Pooling | MAD | CGCNN | de-CGCNN | ALIGNN | de-ALIGNN |
|---|---|---|---|---|---|---|---|
| $U$ | KJ/mol-cell | Average | 25.7 | 17.2 (0.670) | 1.54 (**0.060**) | 13.7 (0.534) | 2.62 (0.101) |
| $U$ | KJ/mol-atom | Average | 5.11 | 3.05 (0.602) | 0.301 (**0.059**) | 3.00 (0.587) | 0.613 (0.120) |
| $U$ | KJ/mol-cell | Sum | 25.7 | 6.53 (0.254) | 1.53 (**0.060**) | 6.88 (0.268) | 1.87 (0.073) |
| $C_v$ | J/(mol-cell*K) | Average | 55.6 | 42.2 (0.759) | 3.25 (**0.058**) | 35.2 (0.632) | 4.95 (0.089) |
| $C_v$ | J/(mol-atom*K) | Average | 13.8 | 8.30 (0.601) | 0.792 (**0.057**) | 8.37 (0.606) | 1.52 (0.110) |
| $C_v$ | J/(mol-cell*K) | Sum | 55.6 | 15.1 (0.272) | 2.86 (**0.051**) | 18.9 (0.339) | 3.75 (0.067) |
| $M$ | μB/formula | Average | 3.13 | 1.29 (0.412) | 1.07 (0.340) | 1.06 (0.340) | 0.964 (**0.308**) |
| $M$ | μB/atom | Average | 0.260 | 0.088 (0.339) | 0.069 (**0.266**) | 0.078 (0.300) | 0.077 (0.296) |
| $M$ | μB/cell | Sum | 6.50 | 1.94 (0.299) | 1.73 (0.267) | 1.63 (**0.251**) | 1.98 (0.304) |

In Table 1, we observe that both de-CGCNN and de-ALIGNN have very significant improvements for the prediction of $U$ and $C_v$. To understand these results, we show the feature importance spectrum of de-CGCNN for prediction of $C_v$ in Figure 4a. We can see that, the human-designed descriptors play important roles in learning $C_v$, with $a$ being the most important feature, while the learned features are much less important. Therefore, the poor prediction ability of CGCNN and ALIGNN for $U$ and $C_v$ can be explained by the fact that $a$ is important to the two properties but CGCNN and ALIGNN cannot learn $a$ well as above. The distribution of feature importance agrees well with the phenomenon in Table 1 that, using the machine learning model based on only human-designed descriptors can have lower errors for prediction of $U$ and $C_v$ compared with the GNNs.

The importance of the input human-designed descriptors to $U$ and $C_v$ can be justified physically as below. Approximately, if we only consider the acoustic phonons (collective vibrations for all atoms in the primitive cell), according to the Debye model of density of states, the phonon internal energy ($U$) and heat capacity ($C_v$) per primitive cell can be written as[58]:

$$U \approx 9k_B T \left(\frac{T}{\theta}\right)^3 \int_0^{x_D} dx \frac{x^3}{e^x - 1} \quad \ldots\ldots(15),$$

$$C_v = \left(\frac{\partial U}{\partial T}\right)_v \approx 9k_B \left(\frac{T}{\theta}\right)^3 \int_0^{x_D} dx \frac{x^4 e^x}{(e^x - 1)^2} \quad \ldots\ldots(16),$$

$$x_D \equiv \frac{\theta}{T} \quad \ldots\ldots(17),$$

$$\theta = \frac{\hbar v}{k_B} \left(\frac{6\pi^2}{V}\right)^{\frac{1}{3}} \quad \ldots\ldots(18),$$

where $\theta$ is the debye temperature, $V$ is the volume of the primitive cell, and $v$ is the velocity of sound, which can be approximated by the first-order Hooke's law:

$$v \approx \sqrt{\frac{C}{m}}d \quad\quad (19),$$

where $C$ is the effective spring constant, $m$ is the mass of atoms in the primitive cell, and $d$ is the effective distance between atomic planes along the direction of vibration. Therefore, with the information of $V$, $C$, $m$, and $d$, we can estimate acoustic $U$ and $C_v$ per primitive cell at given $T$ within the Debye model. Since the set of descriptors in this work includes density and lattice constants, the information of $V$, $m$, and $d$ can be directly obtained by machine learning models from the input descriptors. For $C$, because it is related to the bonding strength, it can be estimated by the bond length-related descriptors. Consequently, machine learning models based on the set of descriptors in this work can approximate $U$ and $C_v$ well within the Debye model, which explains why machine learning model based on only descriptors outperforms CGCNN and ALIGNN in Table 1, as CGCNN and ALIGNN cannot estimate lattice constants well as in Figure 2b. More discussions about the relation between this work and the prediction of $U$ in Legrain *et al.*[59], prediction of $C_v$ in Gurunathan *et al.*[60], prediction of phonon density of states by E3NN[27] and Mat2Spec[31], and prediction of phonon properties from various machine learning force fields[61-64] are provided in the Supporting Information.

$\kappa$ and $M$ are another two properties with improvement from both de-CGCNN and de-ALIGNN (around 10% in these cases). It is known that $\kappa$ depends significantly on periodicity[65], and as shown in Figure 4b, some input descriptors, including $b$, are important to the prediction of $\kappa$. As for $M$, as shown in Figure 4c, some descriptors like structural complexity and lattice constants contribute to the prediction of $M$. In Figure 4d, we show the ratio of feature importance from the human-designed descriptors to the total feature importance from de-CGCNN. We can see that most properties without significant improvement in Table 1 have low contributions from the input human-designed descriptors, with the exception of $v$ and $\|e\|_\infty$ where all the models perform poorly. Here, we choose band gap and total energy for more detailed discussion. For band gap, from Rajan et al.[66] we know that band gap is strongly correlated with volume per atom, which is reasonable as volume per atom reflects the degree of orbital overlap between atoms, and from Wu et al.[67], we know that band gap is strongly correlated with bond angle, which is also reasonable as bond angle represents how atomic orbitals hybridize with each other (such as the difference between $sp^3$ and $sp^2$ hybridization). Since the important features of band gap, such as volume per atom and bond angle, can be well captured by the original GNNs as shown in Figure 2, it is reasonable that the descriptors-hybridized GNNs cannot improve predictions of band gap as the descriptors do not provide useful information that the original GNNs cannot capture. As for total energy, we know that for 3D inorganic materials, short-range interactions dominate the total energy (this is straightforward in structures with covalent bonds; even for ionic systems or metallic systems, the electrostatic screening effect[68] diminishes the role of long-range interaction on total energy). Since the descriptors about short-range interactions can be well captured by the original GNNs as in Figure 2, it is also reasonable that the descriptors-hybridized GNNs cannot improve predictions of total energy. The observation that hybridization with descriptors has larger improvement for CGCNN than ALIGNN might be explained by the fact that, CGCNN captures these descriptors worse than ALIGNN as in Figure 2, therefore hybridization with descriptors provides more missing information to CGCNN than ALIGNN.

Note that, for prediction of extensive materials properties, we have multiple choices for the setting of the regression: average pooling for property normalized per cell/formula, average pooling for property normalized per atom, and sum pooling for property not normalized and scaled with the number of atoms. In order to study the impact of these settings, and how hybridization with descriptors affects the regression with these different settings, we conduct further experiments about these settings on CGCNN, ALIGNN and their hybridized versions for $U$, $C_v$ and $M$. The results are listed in Table 2. Some notable trends are summarized below:

1) For predictions of $U$, $C_v$ and $M$ from original CGCNN and ALIGNN, compared with average pooling for property normalized per cell/formula, the improvement from sum pooling for not normalized property is more significant than average pooling for property normalized per atom. The reason might be the fact that, although the setting of per-atom normalization gets rid of the additional task of predicting number of atoms in the cell, average pooling is less capable to capture the periodicity of the structure than sum pooling, as suggested in Figure 3.

2) When hybridized with descriptors, the difference between the three settings becomes smaller compared with that of the original GNNs. This is also reasonable, as when hybridized with descriptors, the information of number of atoms and periodicity is provided by descriptors, which partially replaces the role of per-atom normalization and sum pooling as discussed in 1).

3) Except de-ALIGNN for $M$ with sum pooling, in all other cases the descriptor-hybridized GNNs outperform the original GNNs. A possible reason for the irregular case is that, as in Figure 4c, when predicting $M$ the importance of descriptors is less than that of the automatically learned features from the GNNs. Therefore, for ALIGNN with sum pooling which already achieves a low predicting error, the positive impact of hybridizing with descriptors might be less significant than the negative impact of information redundancy as discussed below and in Figure S6 in the Supporting Information. Note that here we do not perform descriptors-selection, and we will investigate the impact of descriptors-selection on prediction performance in a future study.

In addition to providing missing information, hybridization with descriptors might have other impacts on the GNNs. In the Supporting Information, we show that the hybridization of descriptors can bias the learned representations less correlated with the input descriptors, although how such bias affects prediction performance is not clear yet. Other questions worth further investigation include: 1) how the improvement scales with dataset size, 2) how to choose the set of input descriptors for optimal performance. It will also be important to understand if the two mentioned behaviors (scaling and selection of descriptors) are similar with or different from that of the descriptors-only models and pure deep representation learning model, and 3) how small geometric distortions, which might lead to the instability of some descriptors, such as coordination number, affect the predictions of descriptors-hybridized deep representation learning, as well as the proposed scheme of learning and predicting descriptors to exam the ability of deep representation learning.

## Discussions and Conclusions

In summary, we propose a systematic approach to analyze the representation power of GNNs for crystal structures. We use human-designed descriptors as a bank of knowledge to test whether CGCNN and ALIGNN can capture knowledge of crystal structures behind descriptors. We find that both GNNs can capture basic local structural descriptors well, but cannot capture the periodicity of crystal structures. We analyze the limitations of the GNNs on capturing periodicity from three perspectives: local expressive power, long-range information and pooling function. We also test the idea of hybridization with descriptors to improve the performance of GNN, and show that descriptors-hybridized CGCNN and ALIGNN have better prediction performance for some materials properties than the original CGCNN and ALIGNN, especially phonon internal energy and phonon heat capacity with 90% lower errors.

The analysis performed in this work can be easily extended to other deep representation learning models, human-designed descriptors, and systems beyond crystals such as molecules and amorphous materials. This work shows that the fields of human-designed descriptors and deep representation learning can be developed synergically. For new deep representation learning models, their ability in representation of crystal structures can be tested by learning existing human-designed descriptors, and for new descriptors, they can be used to reveal how well the existing deep representation learning models capture the knowledge behind these descriptors, which can also be hybridized with deep representation learning models for improved prediction performance. We hope this work may inspire further development of deep representation learning, human-designed descriptors and hybridized machine learning models for crystal structures and materials science.

## Methods

**Datasets.** In this work, we choose 28 (in Figure 2) human-designed descriptors to test their learnability to CGCNN and ALIGNN, and hybridize 29 descriptors (all descriptors in Table 3, except "space group number") with the two GNNs to test the prediction performance. The list of descriptors is provided in Table 3. The criterion for choosing the descriptors is that they are easy to understand and easy to obtain from crystal structures. Number of atoms and lattice constants of the primitive cell are determined by the Niggli reduction[38] implemented in the Structure class in pymatgen[37], and other descriptors are calculated by Matminer[35]. For the descriptor "standard deviation average bond length" (and similar descriptors), the calculation procedure is first calculating the average bond length for each atom, then calculating the standard deviation for the average bond length of all atoms.

In this work, most real 3D crystal structures (primitive cells) and materials properties are downloaded from the Materials Project database (V2021.03.22)[33], and those for $\kappa$ are from the TEDesignLab database[69]. $U$ and $C_v$ are calculated by the PhononDos class in pymatgen[37] based on the phonon density of states from the Materials Project database[33]. The dataset size for each property, shown in Table S1 in the Supporting Information, depends on the number of structures that have the recorded property in the two materials databases. For machine learning of materials properties in Table 1 and Table 2, we split the datasets into 60%, 20% and 20% as the training, validation and test set.

For the dataset used for testing whether CGCNN and ALIGNN can capture human-designed descriptors of crystal structures, since we know that lattice constants of high-symmetry materials are reported to be more learnable than that of low-symmetry materials[41, 42] based on compositions, we create a subset of the Materials Project database ("MP dataset") by removing some structures randomly based on their space group number:

$$Probability(\text{removed}) = \frac{\text{Space group number}}{\text{Space group number}+15},$$

where 15 is the space group number of the C2/c group, the last space group in the class of monoclinic Bravais lattice. Consequently, we have a dataset with 47,862 crystal structures biased to materials with low symmetry to test whether CGCNN and ALIGNN can learn human-designed descriptors from crystal structures. To facilitate the analysis about failure of CGCNN to capture lattice constants, we create a dataset of random 1-dimensional carbon chains ("1D dataset"). The random 1D chains are created by the pseudo-code provided in Section 9 in the Supporting Information. For classification of crystal system in Figure 2c, all the structures in the Materials Project database are used. For the dataset of (1D, short), the number of atoms is set to be between [2, 9), and for the dataset of (1D, long), the number of atoms is set to be between [37, 51). In total, both datasets have 1,400 data points. For machine learning of human-designed descriptors in Figure 2 and 3, we split the dataset into 80%, 10% and 10% as the training, validation and test set.

Table 3. List of abbreviations of descriptors and properties in this work.

| Abbreviations of descriptors | Full name of descriptors | Abbreviations of properties | Full name of descriptors |
|---|---|---|---|
| MAD_in_rela_bond_len | mean absolute deviation in relative bond length | $\log(\kappa)$ | $\log_{10}$ lattice thermal conductivity |
| max_rela_bond_len | maximum relative bond length | $E_{\text{fin.}}$ | final (total) energy per atom |
| min_rela_bond_len | minimum relative bond length | $U$ | phonon internal energy at 300 K |
| max_neighb_dist_var | maximum neighbor distance variation | $C_{\text{v}}$ | constant volume phonon heat capacity at 300 K |
| min_neighb_dist_var | minimum neighbor distance variation | $K$ | bulk modulus |
| range_neighb_dist_var | range neighbor distance variation | $G$ | shear modulus |
| mean_neighb_dist_var | mean neighbor distance variation | $v$ | poisson ratio |
| dev_neighb_dist_var | standard deviation neighbor distance variation | $E_g$ | band gap |
| mean_avg_bond_len | mean average bond length | $\|e\|_\infty$ | modulus of piezoelectric tensor |
| std_avg_bond_len | standard deviation average bond length | $\varepsilon_e.$ | electronic dielectric constant |
| MAD_in_rela_atom_vol | mean absolute deviation in atomic volume | $\varepsilon_t.$ | total dielectric constant |
| mean_avg_bond_ang | mean average bond angle | $n$ | refractive index |
| mean_CN_Voronoi | mean coordination number determined by the Voronoi method[36] | $M$ | total magnetization per formula |
| std_CN_Voronoi | standard deviation coordination number determined by the Voronoi method[36] | | |
| vpa | volume per atom | | |
| packing_frac | packing fraction | | |
| struct_comp_atom | structural complexity per atom | | |
| struct_comp_cell | structural complexity per primitive cell | | |
| natoms | number of atoms per primitive cell | | |
| $a$ | the largest lattice length of the primitive cell | | |
| $b$ | the second largest lattice length of the primitive cell | | |
| $c$ | the smallest lattice length of the primitive cell | | |
| $\alpha$ | the largest lattice angle of the primitive cell | | |
| $\beta$ | the second largest lattice angle of the primitive cell | | |
| $\gamma$ | the smallest lattice angle of the primitive cell | | |
| crys_sys | crystal system | | |

| | |
|---|---|
| space_group_num | space group number |

**Models.** In this work, we use the default architecture of CGCNN[19] and ALIGNN[22] for learning human-designed descriptors in Figure 2 and 3 unless specifically mentioned. The reason for using the default architectures is that, as in Figure 3g and 3h, although intentionally revising their architectures can improve learning performance for some descriptors, in this work we try to show the representation power and limit of CGCNN and ALIGNN in a setting close to those in real applications. For learning materials properties in Figure 4, hyper-parameter search based on the search spaces in Table S2 and S3 is conducted; All the neural networks are trained for 300 epochs[22] on a Quadro RTX 6000 GPU. For feature importance in Figure 4, since the permutation feature importance of deep neural networks is very expensive to calculate, we estimate the feature importance by extracting the representations in equation (12), then feed the representations into a random forest model to calculate the feature importance.

# Data Availability

All datasets and trained machine learning models in this work are provided at:
https://figshare.com/articles/journal_contribution/Improving_deep_representation_learning_for_crystal_structures_by_learning_and_hybridizing_human-designed_descriptors/19654224

Crystal structures and materials properties from the Materials Project database (V2021.03.22) are downloaded at https://materialsproject.org/.

# Code Availability

All codes used in this work are provided at:
https://figshare.com/articles/journal_contribution/Improving_deep_representation_learning_for_crystal_structures_by_learning_and_hybridizing_human-designed_descriptors/19654224

CGCNN can be obtained at https://github.com/txie-93/cgcnn. ALIGNN can be obtained at https://github.com/usnistgov/alignn.

# Supplementary Information for:

# Examining graph neural networks for inorganic crystalline structures: limitation of capturing periodicity

Sheng Gong[1], Tian Xie[2], Yang Shao-Horn[1,3], Rafael Gomez-Bombarelli[1], and Jeffrey C. Grossman[1,*]

[1]Department of Materials Science and Engineering, Massachusetts Institute of Technology, MA 02139, USA
[2]Microsoft Research, Cambridge, United Kingdom CB1 2FB
[3]Computer Science and Artificial Intelligence Lab, Massachusetts Institute of Technology, MA 02139, USA

## Contents

# 1. Relationship between density, volume per atom and lattice

In Figure S1a, we show the correlation between lattice constants and density and volume per atom ("vpa") of structures in the Materials Project database. We can see that, there is no obvious correlation between lattice parameters and density and vpa. As an example, in Figure S1b, we show the lattice constants of FCC- and HCP-closely packed structures. As we know, for elemental structures, FCC and HCP are the two most densely packed structures with the same density and packing fraction (0.74). However, FCC and HCP have different lattice constants and number of atoms in the Niggli primitive cell. This simple example illustrates that, two crystal structures can have very different lattices while share the same density and vpa if they have similar local coordination environments.

The phenomenon that GNN can learn density and vpa better than lattice constants might be attributed to the fact that, determination of lattice requires more global information than that of density and vpa. For example, to determine the vpa of FCC and HCP structures, information of the first coordination shell alone is enough, while to distinguish and determine the lattice of FCC and HCP structures, information of the second coordination shell is also required.

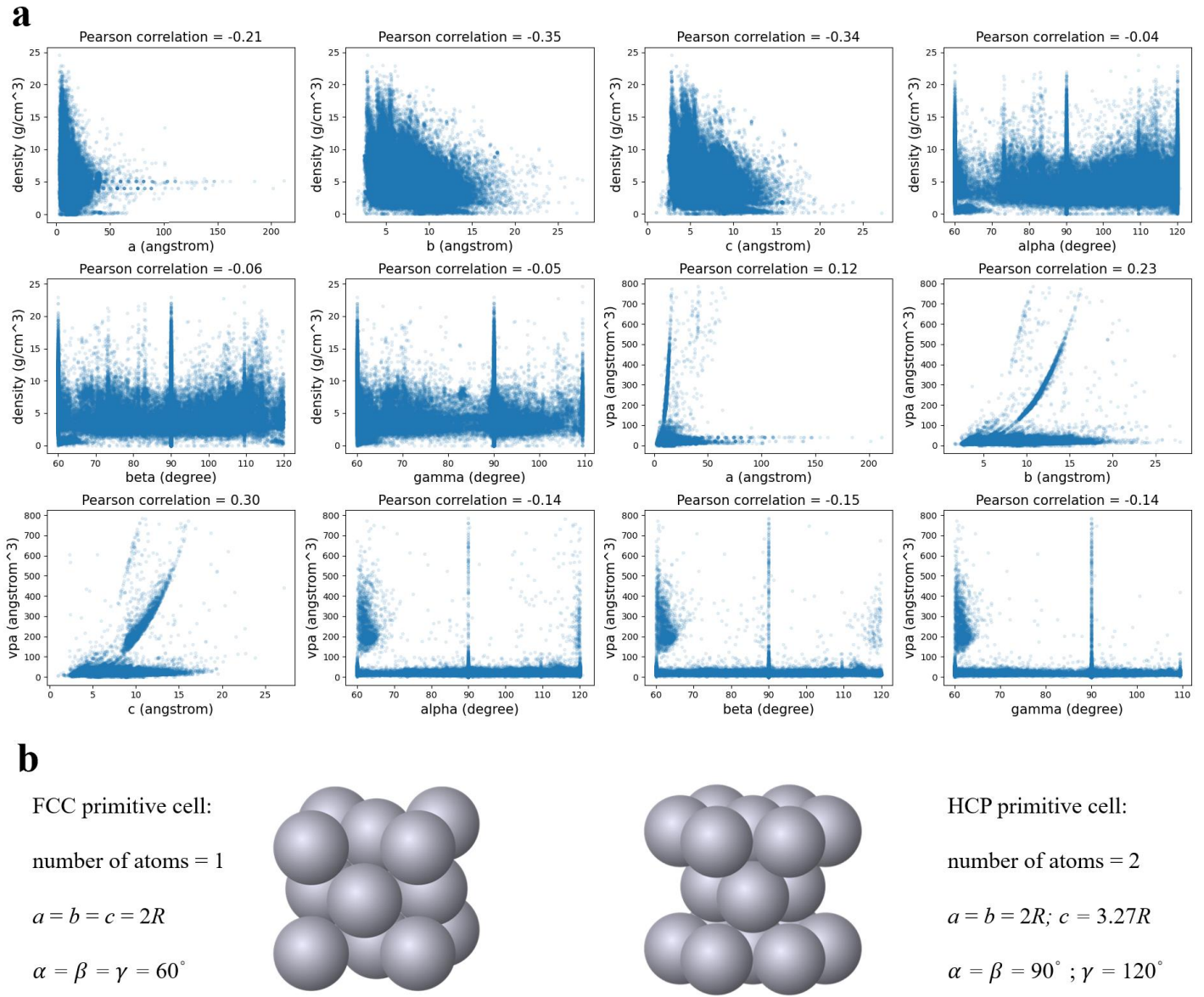

**Figure S1. a** Lattice constants versus density and volume per atom (vpa). **b** Illustration of elemental FCC- and HCP-closely packed structures and their number of atoms and lattice constants of the Niggli primitive cell. *R* is the radius of each atom.

## 2. Limitation of average pooling on extensive properties.

Suggested by the limitation of average pooling on capturing primitive cell-level information, in this section, we show the limitation of average pooling on prediction of extensive properties. Extensive properties scale with number of atoms, such as total energy ($E_{\text{fin.}}$), (phonon) internal energy ($U$),

(phonon) heat capacity ($C_v$) and magnetization ($M$). Despite their extensive nature, in many cases these extensive properties are normalized to intensive versions for data storage and evaluation[16, 18, 19, 22, 32, 33, 70-72], such as energy per atom and magnetization per unit formula. Since intensive properties cannot be learned by sum pooling because of their size invariance[16], for convenience, many GNNs, such as CGCNN and ALIGNN, only implement average pooling and learn the intensive versions of these extensive properties by average pooling in equation (4). Another way to learn extensive properties is illustrated in the bottom branch in Figure S2a, where extensive properties are first learned by machine learning models with sum pooling in equation (13), then normalized for evaluation.

Do the two approaches for learning extensive properties in Figure S2a have the same predictive power? Intuitively, they seem to be equivalent. However, as suggested in Figure 3c and 3h, we know that sum pooling is more powerful to distinguish graphs, and sum pooling can capture periodicity better. This insight suggests that sum pooling might be more powerful than average pooling for learning extensive properties. As shown in Figure S2b, we show the MAE/MAD scores of predictions of four extensive properties by CGCNN and ALIGNN with average pooling and sum pooling. We can see that, sum pooling outperforms average pooling for $U$, $C_v$ and $M$, and has similar performance for $E_{\text{fin.}}$, which verifies our hypothesis that sum pooling might be more powerful than average pooling for learning extensive properties.

Currently, many GNNs designed for prediction of materials properties, such as CGCNN and ALIGNN, only implement size-invariant average pooling function, with the exception of MEGNet where users can easily switch between sum pooling and average pooling. Based on the results shown in this section, we argue that GNN models designed for prediction of materials properties should provide options of pooling functions for different properties, such as average pooling for intensive properties and sum pooling for extensive properties.

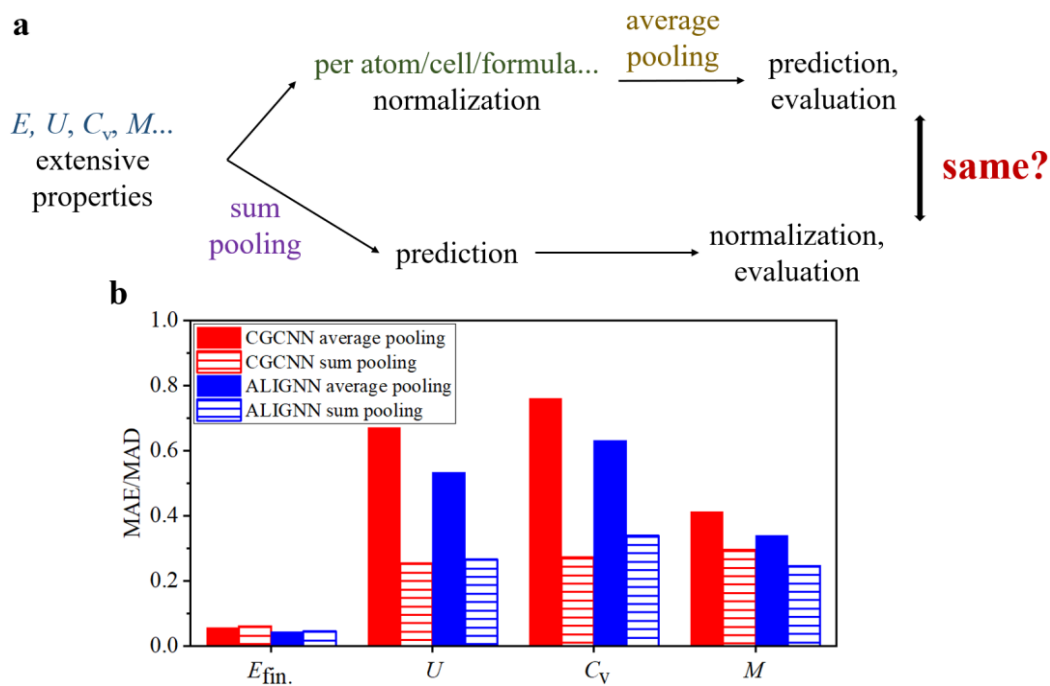

**Figure S2. Average pooling and sum pooling for extensive properties. a** Schematic of two approaches to learn, predict and evaluate prediction performance of extensive properties. **b** MAE/MAD of prediction of four extensive properties by CGCNN and ALIGNN with average pooling and sum pooling.

## 3. Limitation of sum pooling on capturing periodicty.

As mentioned in the main text, even if only primitive cells are input to the GNNs, sum pooling might also fail to capture periodicity in some cases, as periodicity does not always scale with number of atoms in the primitive cells. For example, in Figure 3d we show the case of 1D double chains. Compared with 1D single chains in Figure 3b and 3c, 1D double chains can have similar periodicity but twice number of atoms. In Figure S3, we show the $R^2$ scores of predictions of $a$ of the datasets with 1D single short chains, 1D single and double short chains, 1D single long chains, and 1D single and double long chains, from CGCNN with average pooling and sum pooling, respectively. For the datasets with only single chains, sum pooling outperforms average pooling significantly, while for the datasets with single and double chains, sum pooling can only have very limited improvement over average pooling, which shows the co-existence of single and double chains makes sum pooling harder to determine periodicity than the case with only single chains.

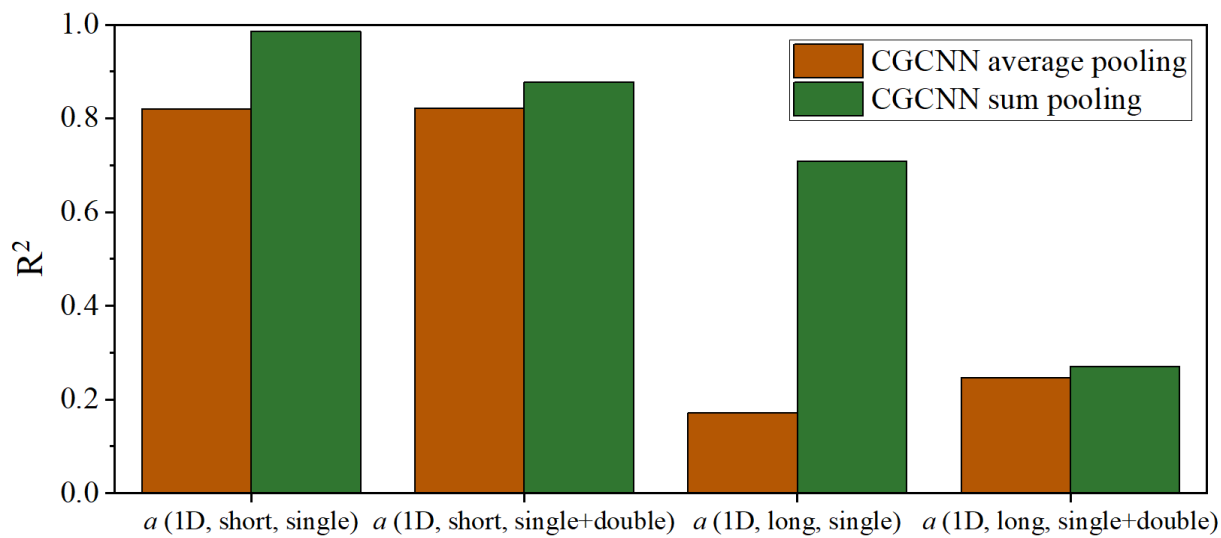

**Figure S3.** $R^2$ scores of predictions of $a$ of the datasets with 1D single short chains, 1D single and double short chains, 1D single long chains, and 1D single and double long chains, from CGCNN with average pooling and sum pooling, respectively.

## 4. Niggli reduced primitive cell and its impact on learning performance of GNN.

As stated in the main text, we use the Niggli reduction to determine the unique Niggli-primitive cell for any given periodic structure[38]. In crystallography, we can use a set of six parameters to define a primitive cell: $(a, b, c, \alpha, \beta, \gamma)$, where the first three parameters are lengths of lattice vectors, and the later three parameters are angles between lattice vectors. In Niggli reduction, we require the lengths of lattice vectors to satisfy the following requirement:

$$a + b + c = \text{minimum}......(S1)$$

If the order of the three lengths are defined, as in the main text we define $a$ the largest one and $c$

the smallest one, then the set of (*a*, *b*, *c*) is unique for any given structure[39]. In other words, the Niggli reduction requires that the primitive cell is composed by three shortest non-coplanar vectors. As shown in Figure S4a, in the example of 2-dimensional square lattice, the Niggli reduced cell is cell 1 (cell 2 and 3 are "primitive cells", but not "Niggli reduced primitive cell"). The lengths of the shortest lattice vectors do have their specific physical meaning. For example, in terms of lattice vibration (acoustic phonon), the lengths of the shortest lattice vectors determine the normal modes of the vibration, such as the normal modes of lattice vibrations of 2-dimensional square lattice in equation S2[58]:

$$\omega^2 = \frac{4K}{M}\sin^2\left(\frac{k_x a}{2}\right) + \frac{4K}{M}\sin^2\left(\frac{k_y a}{2}\right) \ldots\ldots(S2),$$

where $\omega$ is the angular velocity of the normal modes, $K$ is the spring constant, $M$ is the mass of the primitive cell, and $k_x$ and $k_y$ are the quantum numbers of vibrations at the two dimensions, respectively. We can see that the normal modes are separated by $\frac{a}{2}$, not $\frac{\sqrt{2}a}{2}$ or lengths of other possible lattice vectors, which shows that the lattice lengths of the Niggli reduced primitive cell do have special physical meaning that lengths of other possible lattice vectors do not have. Therefore, our observation that GNN cannot capture lattice parameters of Niggli primitive cell is physically meaningful, which naturally leads to the fact that GNN cannot accurately predict properties related to lattice vibration as in the second half of our paper.

However, the set of ($\alpha$, $\beta$, $\gamma$) might not be unique even if (*a*, *b*, *c*) is unique. As mentioned in the main text, both CGCNN and ALIGNN cannot learn the lattice angles well. In Figure S4b, we show the prediction performance of CGCNN, CGCNN with more convolutions, CGCNN with larger limits of number of neighbors and neighboring cut-off radius, CGCNN with sum pooling and ALIGNN for three lattice angles. We can see that both CGCNN and ALIGNN cannot learn the three lattice angles well, and the modifications that improve learning performance of lattice lengths as in the main text do not improve that of lattice angles. The artificial choice of lattice angles might cause the poor learning performance. For example, in Figure S4c, we plot the structure of simple hexagonal structure. We can see that the three lengths of lattice vectors are unique and reflect the intrinsic characteristics of the structure, such as the three minimal distances between the smallest repeating units in three dimensions. For ($\alpha$, $\beta$, $\gamma$), we know that two of them are 90° (angles between the vertical lattice vector and the two in-plane lattice vectors), but there are actually two choices of the third one: 60° and 120°, and two choices of ($\alpha$, $\beta$, $\gamma$): (120°, 90°, 90°) and (90°, 90°, 60°). Although in Niggli reduction, (120°, 90°, 90°) is chosen as the lattice angles, this choice is artificial to the given structure without clear physical meaning, to the best of the authors' knowledge. Therefore, machine learning algorithms might not be able to capture the artificially determined characteristics of crystal structures. Further studies are necessary to design more intrinsic description of relative orientation between the lattice vectors.

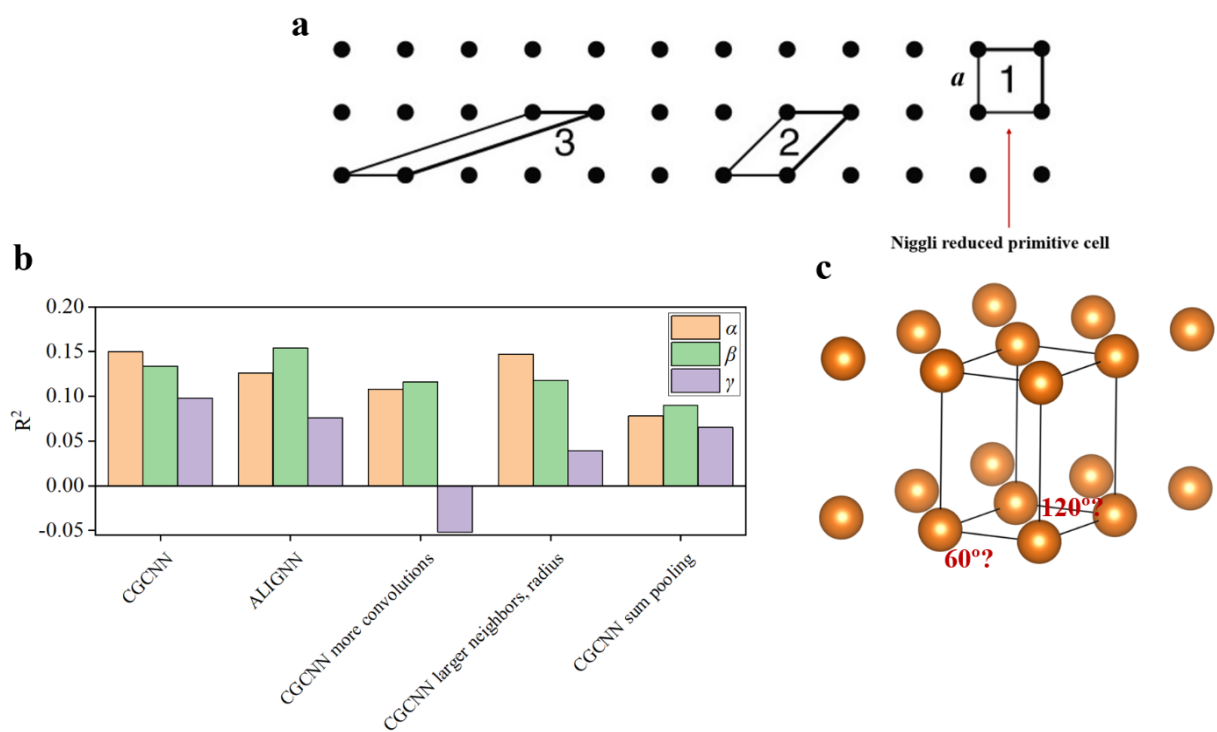

**Figure S4. a** $R^2$ scores of prediction of lattice angles from CGCNN, ALIGNN and variants of CGCNN. Alpha denotes the largest lattice angle, and gamma the smallest one. **b** Illustrations of ambiguity of choice of lattice angles by the example of simple hexagonal primitive cell.

## 5. Receptive field of GNNs for inorganic crystal structures.

In order to estimate the length of the receptive field of GNNs in real 3D crystal structures, we estimate the average length of the receptive field by equations S3 and S4:

$$V = \frac{4}{3}\pi r^3 \approx n * vpa \quad \text{......(S3)},$$

$$r = \begin{cases} \sqrt[3]{\frac{3}{4\pi} * n * vpa}, & \text{if } r < r_{cut} \\ r_{cut}, & \text{if } r \geq r_{cut} \end{cases} \quad \text{......(S4)},$$

where $V$ is the volume within the sphere, $n$ is the number of neighbors, $vpa$ is the volume per atom of the crystal structure, $r$ is the radius of the sphere (half of the length of the receptive field in one convolution step in GNNs), and $r_{cut}$ is the cut-off radius of the GNNs (8 Å for default CGCNN and ALIGNN). In Figure S5a, we show the distribution of $r$ with $n = 12$ (default maximum number of neighbors in CGCNN and ALIGNN), from which we can see that the average $r$ is around 3.4 Å, therefore with three steps of convolutions the average length of receptive fields is around 20 Å.

As shown in Figure S5b, although most crystal structures in the Materials Project database have lattice around 10 Å, there are still some structures with long periodicity. For example, there are 6,000 structures with the longest lattice vector longer than 20 Å. Two examples of the long structures are provided in Figure S5c, which are $Mo_9O_{25}$ (mp-28777, $a = 28.4$ Å, energy above hull = 0.008 eV/atom) and $Pr_2Au_5F_{21}$ (mp-14715, $a = 26.3$ Å, energy above hull = 0 eV/atom). Therefore, our 1D toy dataset and analysis is still relevant to actual usages of GNNs as they represent thousands of materials where the receptive fields of typical GNNs are shorter than the periodicity of the crystal structures.

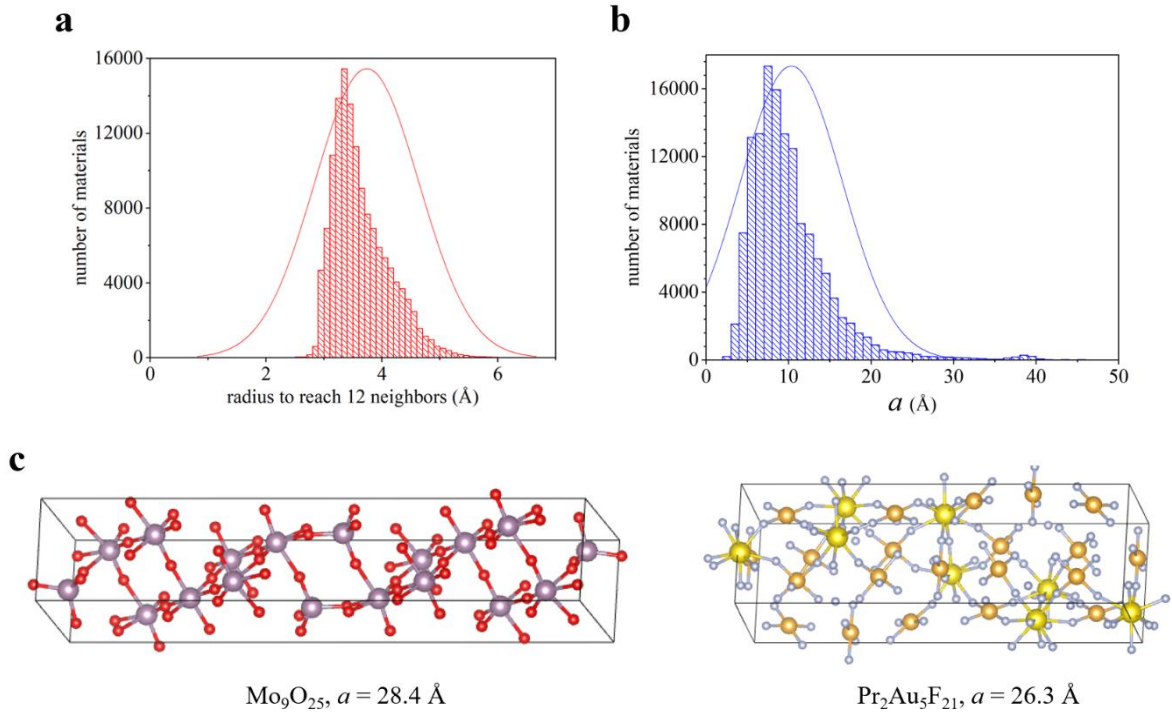

**Figure S5. a** Distribution of the average radius to reach 12 neighbors of all structures in the Materials Project database. **b** Distribution of $a$ of all structures in the Materials Project database. **c** Two examples of structures with long periodicity. Color coding: red: O; purple: Mo; gold: Pr; orange: Au; grey: F.

## 6. Structural Complexity

Structural complexity is defined as below:

$$S = -N \sum_{i=1}^{k} \frac{m_i}{N} \log_2 \frac{m_i}{N} \dots\dots(S2),$$

where $N$ is the total number of atoms in the primitive cell, $k$ is the number of symmetrically inequivalent sites, $m_i$ is the number of sites classified in the $i^{\text{th}}$ symmetrically inequivalent site. Structural complexity quantifies the complexity of sites distribution in a structure, as larger complexity, more different symmetrically inequivalent sites in the structure. Structural complexity per primitive cell is calculated as equation S2, and structural complexity per atom is calculated by equation S2 divided by the number of atoms in the primitive cell.

## 7. More discussions about $U$ and $C_v$

Recently, Legrain et al.[59] reported machine learning of $U$ by compositions of materials. The main difference between this work and Legrain et al.[59] is that, in Legrain et al.[59], only 292 materials are included in the dataset, while in this work about 1,500 materials are included in the datsets for $U$ and $C_v$. On the other hand, Mat2Spec[31] and E3NN[27] are proposed to predict phonon density of states, and consequently, heat capacity. Unfortunately, we cannot easily compare our predictions and predictions from the two models mentioned above, as the $C_v$ in this work is based on full phonon density of states, while Mat2Spec[31] and E3NN[27] are designed to predict filtered and truncated phonon density of states with 51 frequencies up to 1000 cm$^{-1}$. The reason for the success of E3NN for predicting $C_v$ might be that E3NN employs equivariant representations[27] with high local expressive power, and the reason for the success of Mat2Spec for predicting $C_v$ might be the explicit exploitation of correlations of density of states between frequencies in phonon density of states. Very recently, Gurunathan et al.[60] employ ALIGNN to predict phonon density of states, and find that $C_v$ from predictions of phonon density of states is more accurate than $C_v$ from direct prediction, which provides further insights into predictions of phonon related properties.

On the other hand, phonon internal energy and heat capacity can also be predicted by machine learning force fields, such as Ladygin et al. where a moment tensor potential is used to study phonon properties of Al, Mo, Ti, and U[61], Babaei et al. where a Gaussian approximation potential is used to study phonon properties of Si[64], and Dhaliwal et al. where a potential based on random Fourier features is used to study phonon properties of graphene[63]. However, there still lacks a universal machine learning force field that has shown the ability to calculate phonon properties for different materials with higher accuracy than direct prediction of phonon properties by data-driven machine learning models. One of the most recent universal machine learning force field is Chen et al.[62], where a GNN-based universal machine learning force field (UMLFF) is trained on the trajectories of geometry relaxation of structures in the Materials Project database[62]. The UMLFF is shown to have accurate predictions of energy ($R^2 = 0.959$) and force ($R^2 = 0.984$) for different inorganic materials with almost all elements in the periodic table. In order to directly compare the performance of predicting phonon-related properties by our descriptors-hybridized graph neural networks (de-CGCNN and de-ALIGNN) and the UMLFF, we use the UMLFF to calculate phonon internal energy ($U$) and heat capacity ($C_v$) at 300K of materials in the same test set as de-CGCNN and de-ALIGNN. The MAE/MAD ratio of $U$ and $C_v$ by the UMLFF are 0.1115 and 0.1298, respectively. According to Table 2 in the main text, the MAE/MAD ratio of $U$ and $C_v$ by de-CGCNN can be as low as 0.059 and 0.051, respectively, which demonstrates the importance of this work as it provides more accurate predictions of phonon-related properties than the current UMLFF. The lower error of de-CGCNN compared with the UMLFF for $U$ and $C_v$ might result from the fact that, the de-CGCNN models are specifically trained on the target properties ($U$ and $C_v$), while the UMLFF is trained on energies and forces of geometric configurations near equilibrium and some of the distorted configurations important for phonon calculations might not be covered in the training set.

## 8. Impact of hybridization with descriptors on the learned representations

As in equation (14) in the main text, since the descriptors participate in the optimization of deep representation learning, the inclusion of descriptors would affect the optimization of the learned representations. In other words, consider the gradient propagation in the optimization of the representation learning:

$$\frac{\partial L}{\partial w_{pq}^R} = \frac{\partial z_q^{R+1}}{\partial w_{pq}^R} * \frac{\partial L}{\partial z_q^{R+1}} \ldots\ldots \text{(S3)},$$

where $L$ is the loss function, $w_{pq}^R$ is the weight from the $p^{th}$ unit of the representation layer ($\frac{1}{N_a}\sum_{i=1}^{N_a} a_i^{n^*} \oplus$ descriptors) in equation (14) to the $q^{th}$ unit of the layer after the representation layer (first layer of FCN in equation (14)), and $z_q^{R+1}$ is the $q^{th}$ unit of the layer after the representation layer.

Therefore, after inclusion of descriptors, $z_q^{R+1}$ changes, and consequently $\frac{\partial L}{\partial w_{pq}^R}$ changes even if the $p^{\text{th}}$ unit is from the part $\frac{1}{N_a}\sum_{i=1}^{N_a} a_i^{n^*}$, and all gradients before the representation layer change due to the chain rule of gradient propagation.

How does the change of gradient affect the learned representations? As an example, in Figure 4c, we show the feature importance spectrum of de-CGCNN for prediction of $\kappa$, from which we can see that the learned representations play the most important roles, and some descriptors contribute significantly such as mean and standard deviation of bond length, volume per atom and $b$. Except $b$, the other three important descriptors can be well learned by CGCNN as shown in Figure 2a. In order to understand this phenomenon, we investigate how well the learned representations from CGCNN correlate with descriptors before and after the hybridization with descriptors. In other words, we investigate the following correlation:

$$\text{Corr}(\frac{1}{N_a}\sum_{i=1}^{N_a} a_i^{n^*}, \text{descriptors})\ldots\ldots(S4).$$

In Figure S6a, we show the correlation between the learned representations and each descriptor for learning $\kappa$, and we can see that the learned representations from de-CGCNN are less correlated with the descriptors than that from CGCNN. The weaker correlation after the inclusion of descriptors supports our hypothesis that hybridization with descriptors pushes the optimization of learned representations away from the already known information in the input human-designed descriptors.

As a comparison, we construct machine learning models based on learned representations from CGCNN and human-designed descriptors as below:

$$\text{Output} = \text{FCN}(\frac{1}{N_a}\sum_{i=1}^{N_a} a_i^{n^*}(\text{already learned from CGCNN}) \oplus \text{descriptors}) \ldots\ldots (S5).$$

The main difference between equation (S5) (named as "CGCNN+descriptors") and equation (14) (de-CGCNN) is that, descriptors participate in the optimization of learned representations in de-CGCNN, while descriptors do not in CGCNN+descriptors. In Figure S6b, we show the MAE/MAD ratio of machine learning models based only on descriptors, CGCNN, CGCNN+descriptors, and de-CGCNN. We can see that, for most properties, de-CGCNN has lower error than CGCNN+descriptors, which supports the proposed mechanism that participation of descriptors in the optimization of representations improves the performance of deep representation learning. For $\kappa$, $U$, $C_v$, and $M$, we observe that CGCNN+descriptors outperforms CGCNN, and such improvement mainly comes from the missing information in descriptors as discussed in the main text. For most remaining properties, CGCNN+descriptors has similar performance with CGCNN, which suggests that the improvement of de-CGCNN for these properties might come from the participation of descriptors in the optimization of representations.

Despite the observed improvement, intuitively, hybridization with descriptors in equation (14) and addition of descriptors in equation (S5) would have a negative impact on prediction performance due to the introduction of redundant information from these two modifications[73, 74]. Such redundancy can be seen in Figure S6a, where learned representations from both CGCNN and de-CGCNN are correlated with human-designed descriptors in some degree. In Figure S6b, we observe that CGCNN+descriptors has higher error than the model based on only descriptors for $U$ and $C_v$. Since the only difference between the two models is the presence of learned representations in CGCNN+descriptors, the increase of error associated with CGCNN+descriptors supports the proposed mechanism that redundant information can harm the predictive power of machine learning models[73, 74].

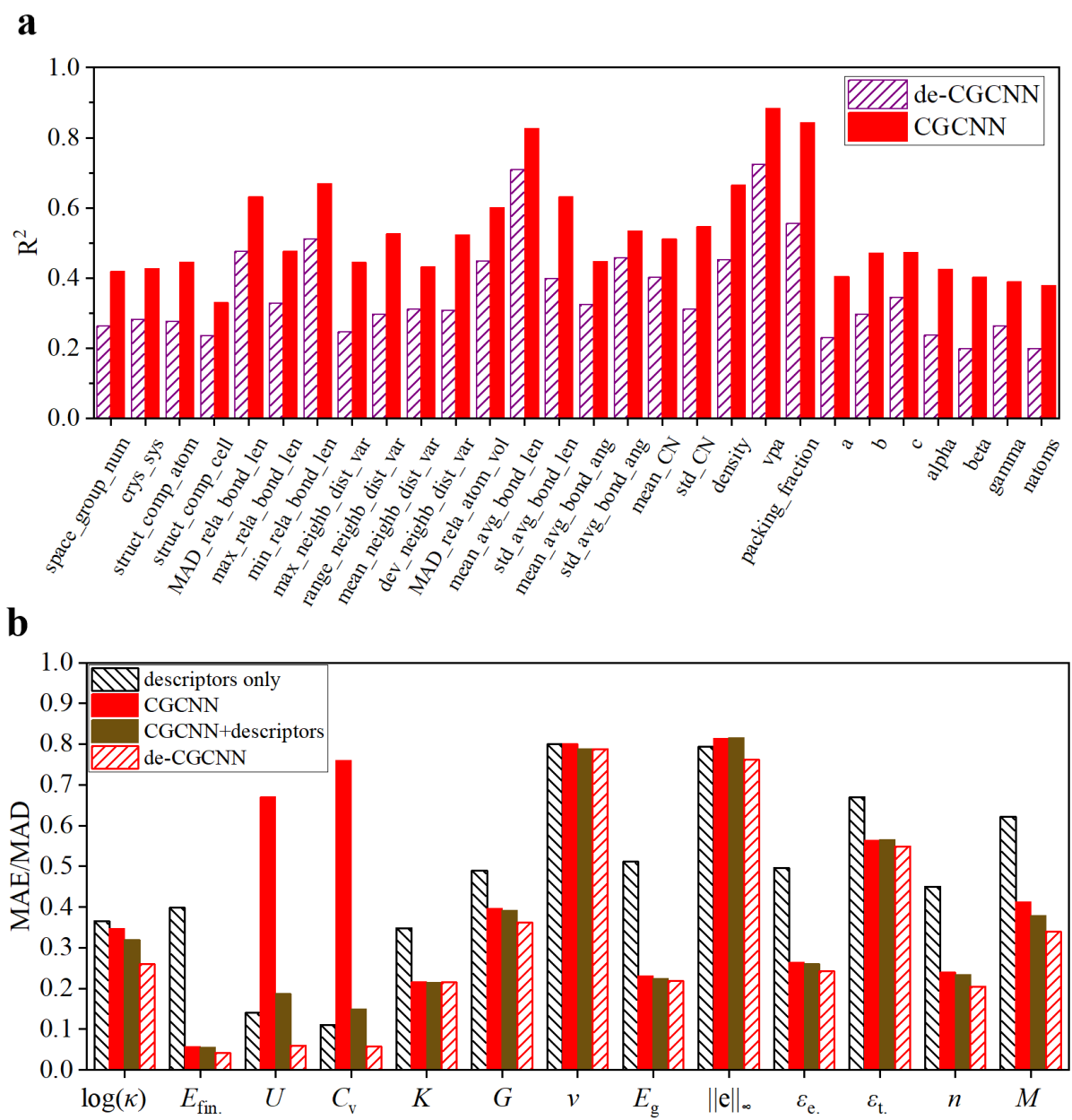

**Figure S6. a** $R^2$ scores of linear regressions between each descriptor and the learned representations from CGCNN and de-CGCNN. **b** MAE/MAD ratio of prediction of 13 materials properties from machine learning models based only on descriptors, CGCNN, machine learning models based on CGCNN-learned representations and descriptors, and de-CGCNN.

# 9. Pseudocode to generate the 1D chains

**For each chain:**

  **pos = []** #initialize positions of atoms

  **for j in range(n):** # number of atoms in the chain

    **if j == 0:**

      **pos.append([3*random, 3*random, 3*random])**

      # first atom, random coordinates in all three dimensions

      # 3 = 2*1.5 Å (1.5 Å to approximate C-C bond length)

      # random: random number between (0, 1)

    **elif j%2 == 0:**

      **pos.append([pos[j-1][0] + 3*random, pos[j-1][1] + 3*random, pos[j-1][2] + 3*random]))**

      # even number of atom, random displacement from the previous

      # atom in all the three dimensions

    **else:**

      **pos.append([pos[j-1][0] + 3*random, pos[j-1][1] - 3*random, pos[j-1][2] - 3*random])**

      # odd number of atom, elongation in the first dimension, retraction in the other

      # two dimensions to keep the chain intact when creating lattice.

**a = pos[-1][0]; b = 100; c = 100**

# position of the last atom as the end of the cell, add vacuum for b and c

**lattice = Lattice.from_parameters(a, b, c, 90, 90, 90)**

## 10. Supplementary Tables

Table S1. Ratio of materials (related to the 140k materials in the Materials Project V2021.03.22) with each property and prediction results of machine learning models for the lattice thermal conductivity in the TEDesignLab database and 12 properties in the Materials Project database. The value in each parenthesis in the bottom of each cell is the MAE/MAD ratio. Here all the GNN models are used with average pooling.

| Property | Unit | Ratio of materials | MAD | MAE of descriptors + RF | MAE of CGCNN +descriptors | MAE of CGCNN | MAE of de-CGCNN | MAE of ALIGNN | MAE of de-ALIGNN |
|---|---|---|---|---|---|---|---|---|---|
| $\log(\kappa)$ | | 0.020 | 0.427 | 0.156 (0.365) | 0.137 (0.320) | 0.148 (0.347) | 0.111 (0.260) | 0.117 (0.276) | 0.106 (0.248) |
| $E_{fin.}$ | eV/atom | 1.0 | 1.29 | 0.513 (0.398) | 0.071 (0.055) | 0.073 (0.057) | 0.054 (0.042) | 0.058 (0.045) | 0.068 (0.052) |
| $U$ | KJ/mol-cell | 0.011 | 25.7 | 3.60 (0.140) | 4.81 (0.187) | 17.2 (0.670) | 1.54 (0.060) | 13.7 (0.534) | 2.62 (0.101) |
| $C_v$ | J/(mol-cell*K) | 0.011 | 55.6 | 6.16 (0.111) | 8.29 (0.149) | 42.2 (0.759) | 3.25 (0.058) | 35.2 (0.632) | 4.95 (0.089) |
| $K$ | GPa | 0.095 | 56.4 | 19.6 (0.348) | 12.2 (0.215) | 12.3 (0.217) | 12.2 (0.215) | 11.5 (0.204) | 11.6 (0.206) |
| $G$ | GPa | 0.095 | 30.6 | 15.0 (0.489) | 12.0 (0.392) | 12.2 (0.397) | 11.1 (0.362) | 10.3 (0.337) | 9.99 (0.326) |
| $v$ | | 0.095 | 0.064 | 0.051 (0.800) | 0.050 (0.788) | 0.051 (0.801) | 0.050 (0.788) | 0.046 (0.716) | 0.045 (0.702) |
| $E_g$ | eV | 1.0 | 1.26 | 0.643 (0.511) | 0.282 (0.224) | 0.290 (0.231) | 0.274 (0.218) | 0.247 (0.196) | 0.249 (0.198) |
| $\|\|e\|\|_\infty$ | $C/m^2$ | 0.024 | 0.555 | 0.440 (0.793) | 0.452 (0.815) | 0.451 (0.813) | 0.423 (0.762) | 0.442 (0.796) | 0.411 (0.742) |
| $\varepsilon_{e.}$ | | 0.052 | 2.82 | 1.40 (0.496) | 0.736 (0.261) | 0.744 (0.264) | 0.683 (0.242) | 0.602 (0.213) | 0.604 (0.214) |
| $\varepsilon_{t.}$ | | 0.052 | 9.06 | 6.06 (0.670) | 5.13 (0.566) | 5.10 (0.563) | 4.97 (0.549) | 4.51 (0.498) | 4.43 (0.489) |
| $n$ | | 0.052 | 0.590 | 0.266 (0.450) | 0.138 (0.234) | 0.142 (0.240) | 0.121 (0.205) | 0.116 (0.197) | 0.114 (0.194) |
| $M$ | μB/formula | 1.0 | 3.13 | 1.95 (0.622) | 1.19 (0.379) | 1.29 (0.412) | 1.07 (0.340) | 1.06 (0.340) | 0.964 (0.308) |

Table S2. Hyper-parameter search space for CGCNN, CGCNN with sum pooling, and de-CGCNN. Parameters not mentioned here are set to the default value as in the open source codes.

| Name | Space |
|---|---|
| atom feature length | 32, 64 |
| hidden feature length | 64, 128 |
| number of hidden layers | 1, 2, 3 |
| learning rate | 1e-3, 1e-2 |

Table S3. Hyper-parameter search space for ALIGNN, ALIGNN with sum pooling, and de-ALIGNN. Parameters not mentioned here are set to the default value as in the open source codes.

| Name | Space |
|---|---|
| edge input feature length | 40, 80 |
| hidden feature length | 64, 128, 256 |
| triplet input feature length | 20, 40 |
| learning rate | 1e-3, 1e-2 |